\documentclass[12pt,a4paper]{article}

\usepackage[english]{babel}
\usepackage[utf8]{inputenc}
\usepackage[T1]{fontenc}
\usepackage{graphicx}
\usepackage[colorlinks=true,citecolor=blue,linkcolor=black,urlcolor=blue,pdfencoding=auto,psdextra]{hyperref}
\usepackage{cite}
\usepackage{amsmath}
\usepackage{amssymb}
\usepackage{bbm}
\usepackage{xcolor}
\usepackage{booktabs}

\usepackage{lineno}
\usepackage{textcomp}

\newcommand{\ket}[1]{\left| #1 \right>} 

\newcommand {\grsim} {\ {\raise-.5ex\hbox{$\buildrel>\over\sim$}}\ }
\newcommand {\lessim} {\ {\raise-.5ex\hbox{$\buildrel<\over\sim$}}\ }

\newcommand{\invisiblesection}[1]{%
  \phantomsection%
  \stepcounter{section}%
  \addcontentsline{toc}{section}{\protect\numberline{\thesection}#1}%
  }

\begin{document}

\title{Fermionic time-reversal symmetry \\ in a photonic topological insulator}

\author{Lukas J. Maczewsky$^{1,\ast}$, Bastian Höckendorf$^{2,\ast}$,\\ Mark Kremer$^{1}$, Tobias Biesenthal$^{1}$, Matthias Heinrich$^{1}$\\ Andreas Alvermann$^{2}$, Holger Fehske$^{2}$, and Alexander Szameit$^{1}$\\
\normalsize{$^1$ Institut f\"ur Physik, Universit\"at Rostock, Albert-Einstein-Str. 23,}\\ \normalsize{18059 Rostock, Germany.}\\
\normalsize{$^2$ Institut für Physik, Universität Greifswald, Felix-Hausdorff-Str. 6,}\\ \normalsize{17489 Greifswald, Germany.}\\
\normalsize{$^\ast$ These authors contributed equally.}\\
\textit{\normalsize{alvermann@physik.uni-greifswald.de; alexander.szameit@uni-rostock.de}}
\\
}

\maketitle


\vspace*{-20pt}

\invisiblesection{Intro}
{\textbf{Much of the recent enthusiasm directed towards topological insulators \cite{Konig2007,Hsieh,HasanKane2010,Lu2014,Soljacic,Rechtsman,Hafezi2013,Yang2015,susstrunk2015,stutzer2018,bandres2018,blanco2018,klembt2018} as a new state of matter is motivated by their hallmark feature of protected chiral edge states. In fermionic systems, Kramers degeneracy gives rise to these entities in the presence of time-reversal symmetry (TRS) \cite{KaneMelePRL, BernevigZhangPRL, Konig2007, Hsieh,HasanKane2010}. In contrast, bosonic systems obeying TRS are generally assumed to be fundamentally precluded from supporting edge states\cite{HasanKane2010,ozawa2018}. In this work, we dispel this perception and experimentally demonstrate counter-propagating chiral states at the edge of a time-reversal-symmetric photonic waveguide structure. The pivotal step in our approach is encoding the effective spin of the propagating states as a degree of freedom of the underlying waveguide lattice, such that our photonic topological insulator is characterised by a $\mathbb{Z}_2$-type invariant. Our findings allow for fermionic properties to be harnessed in bosonic systems, thereby opening new avenues for topological physics in photonics as well as acoustics, mechanics and even matter waves.
}

\pagebreak
With the advent of topological insulators (TIs) \cite{KaneMelePRL, BernevigZhangPRL}, material science began to lift the veil from an entirely new realm of physics. In a seemingly paradoxical fashion, solid-state TIs prohibit electrons from being traversing their interior, while simultaneously supporting chiral surface currents that are protected by particle-number-conservation and time-reversal symmetry (TRS) \cite{Konig2007, Hsieh,HasanKane2010}. Due to the latter mechanism, pairs of counter-propagating states with opposite spin exist (see Fig.\ref{fig1a}(a)), while scattering between them is strongly suppressed. As a result, a TI's surface may be highly conductive while the bulk remains insulating. This phase of a material is characterised by a $\mathbb Z_2$ topological invariant instead of a Chern number \cite{KaneMelePRL, BernevigZhangPRL}, and occurs naturally only in fermionic systems. Bosonic systems, in contrast, do not exhibit Kramers degeneracy, and
are thus not expected to support topologically protected counter-propagating edge states in the presence of TRS.
Instead, topological phases with non-trivial Chern number \cite{haldane1988model, Haldane} and co-propagating edge states
can be induced by means of magnetic fields or external driving that break TRS. Various incarnations of such Chern-type bosonic TIs have been implemented across a broad range of physical platforms, such as microwave systems \cite{Soljacic}, photonic lattices \cite{Rechtsman}, matter waves \cite{jotzu2014}, acoustics \cite{Yang2015}, and even mechanical waves \cite{susstrunk2015}.
Quite generally, bosonic systems are believed to require breaking of TRS in order to elicit topologically non-trivial behaviour.

\vspace*{5pt}
In our work, we challenge this perception and devise as well as experimentally implement a photonic TI with unbroken fermionic TRS. In essence, we judiciously drive a bosonic system to form counter-propagating, scatter-free and topologically protected edge states.
Our driving protocol, which is outlined below and described in more detail in the Supplementary Information,
is designed specifically to impose Kramers degeneracy on the photonic TI.
The core idea behind this approach is to encode the spin degree of freedom of fermionic particles as an effective pseudo-spin degree of freedom of the underlying photonic waveguide lattice, as shown in Fig.\ref{fig1a}(b).
This results in the band structure shown in Fig.\ref{fig1b}, where two counter-propagating chiral edge states appear in the band gap of the bulk as signature of topological protection in the presence of fermionic TRS.

\begin{figure}[ht]
    	\centering
 	\includegraphics[width=0.5\textwidth]{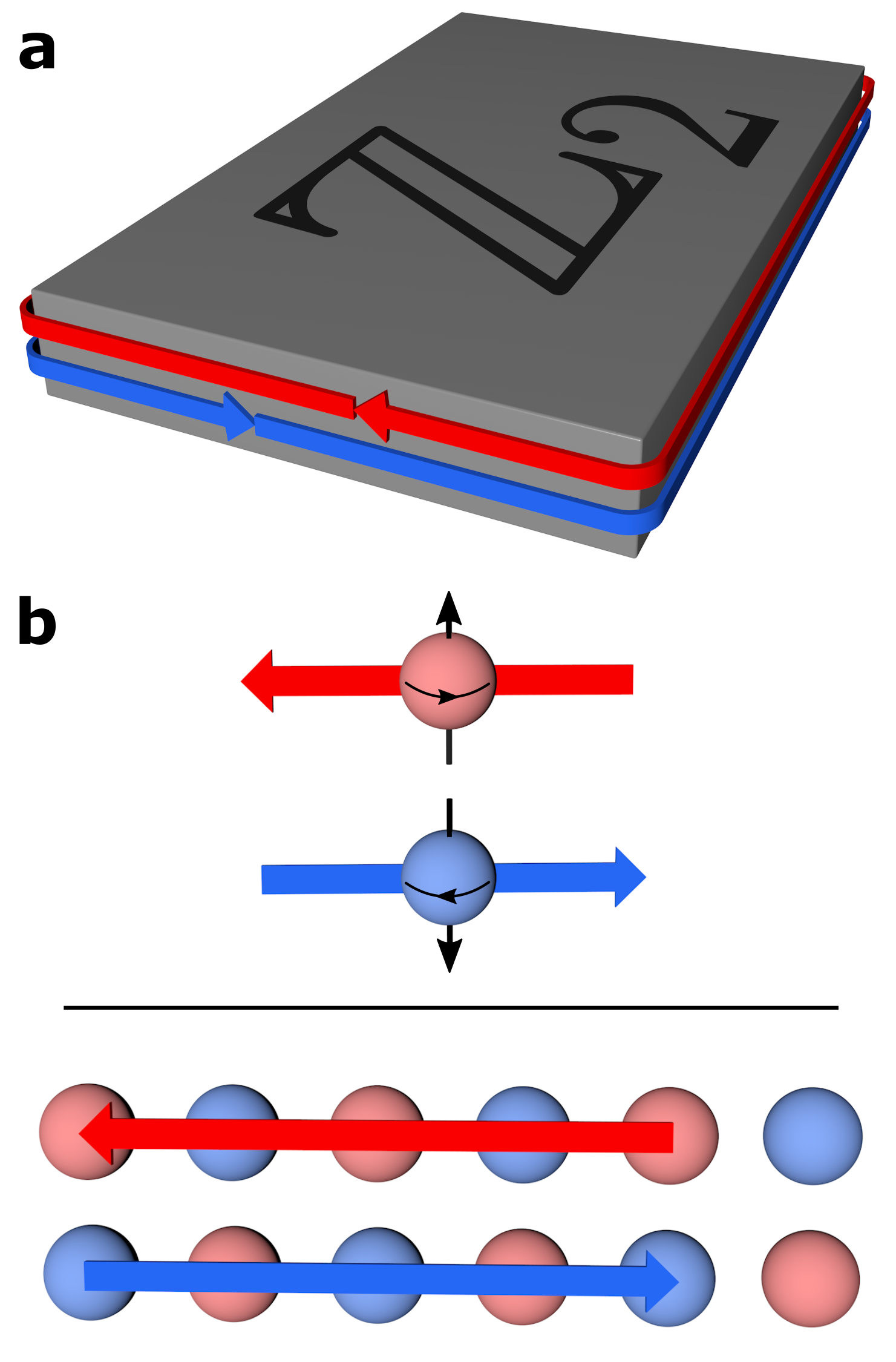}
         \vspace{-8pt}
 	\caption{\label{fig1a} \textbf{Conceptual idea.} (a)~A conventional two-dimensional fermionic topological insulator supports two counter-propagating protected chiral edge states. (b)~In accordance with TRS, these states exhibit opposite spin orientations (upper panel). By mapping the spin property onto a degree of freedom of the underlying lattice, a pseudo-spin can be implemented for bosonic edge states (lower panel), while maintaining the fermionic TRS.
}
\end{figure}

\begin{figure}[ht]
    	\centering
 	\includegraphics[width=0.6\textwidth]{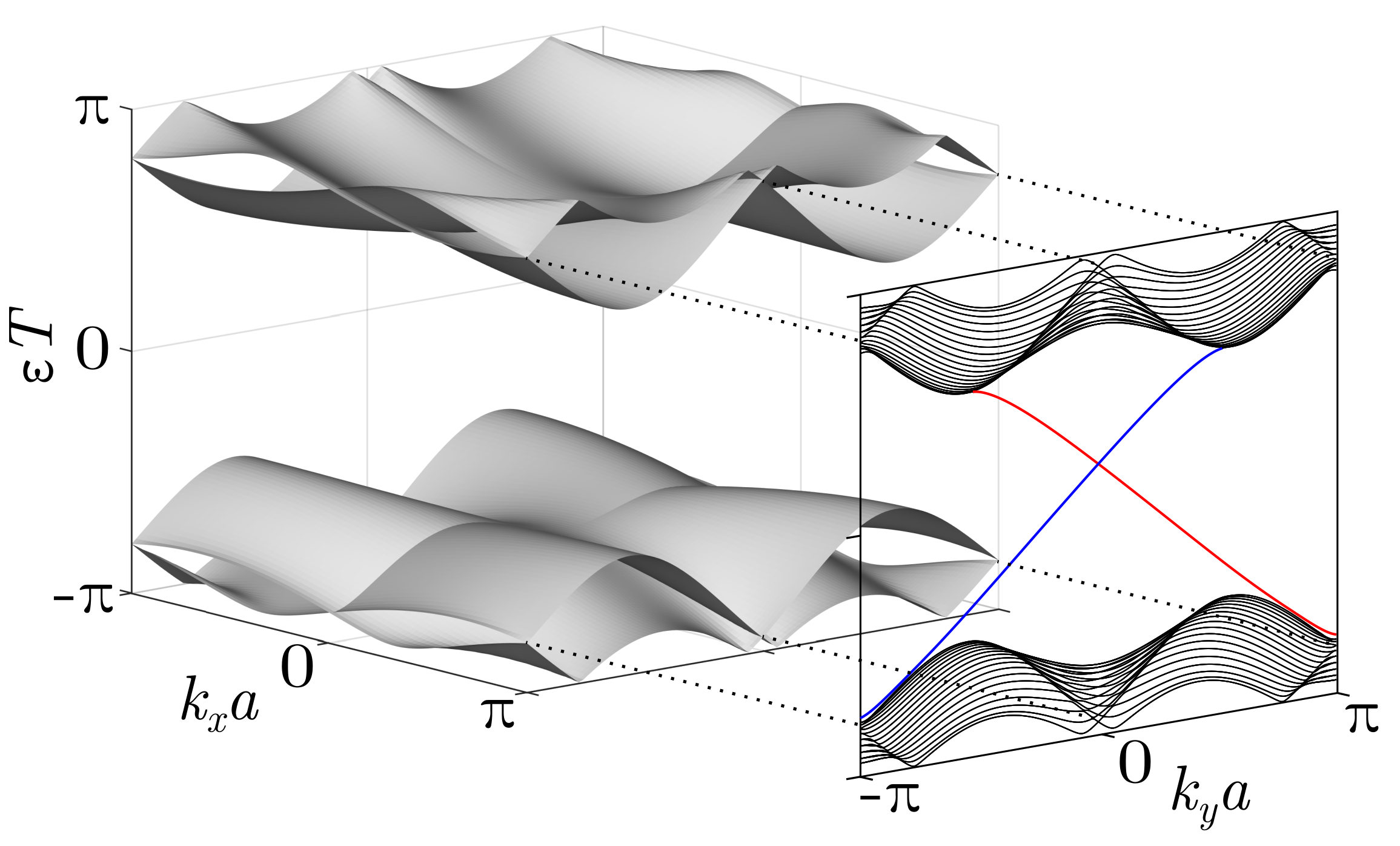}
         \vspace{-8pt}
 	\caption{\label{fig1b} \textbf{Band structure.} The three-dimensional band structure established by our driving protocol is periodic in the quasi-momenta $k_x$, $k_y$ as well as in the quasi-energy $\varepsilon$. Whereas the bulk band structure clearly shows the insulating gap of the system, the band structure of the edge, shown in front, contains the two chiral counter-propagating edge modes (in blue and red, respectively) in the band gap.
}
\end{figure}

Our approach is inspired by the construction of the $\mathbb Z_2$ topological insulator according to Kane/Mele, Bernevig/Zhang, Carpentier \cite{KaneMelePRL, BernevigZhangPRL, Carpentier}, where the combination of two inverse Chern insulators results in a system that is symmetric under time reversal and supports counter-propagating edge states. In our work, we adapt this concept to Floquet systems, and superimpose two inversely driven anomalous topological insulators \cite{Lindner, Maczewsky, Mukherjee} to obtain a $\mathbb{Z}_2$ Floquet TI. The corresponding driving protocol is implemented on two intertwined sublattices (marked with either red (R) and blue (B) sites in Fig.~\ref{fig2}(a)), which represent the two states associated with a fermionic pseudo-spin $1/2$ via the encoding described above. A full driving cycle is comprised of a sequence of six individual steps, each of which couples two different nearest-neighbour sites as indicated by the dotted lines. These steps implement two fundamental types of operations ~\cite{KitagawaPRA, Carpentier}: Steps 1, 3, 4, and 6 realise spin-preserving translations through interactions between sites of the same sublattice. On the other hand, steps 2 and 5 manifest spin rotations by connecting sites from different sublattices. In the latter, partial hopping represents general spin rotations, whereas a spin flip is established by full population exchange between the sublattices.

Figure ~\ref{fig2}(b) illustrates the evolution of single-site excitations in the case of a spin flip. Note how, depending on the initially excited sublattice, the sequence of alternating nearest-neighbour couplings prescribed by the driving protocol gives rise to two distinct edge states, moving either counter-clockwise (red arrow) or clockwise (blue arrow). In contrast, all excitations in the bulk of the lattice follow closed loops such that no effective transport can occur and the wave packets remain localised.

\begin{figure}
    	\centering
 	\includegraphics[width=1\textwidth]{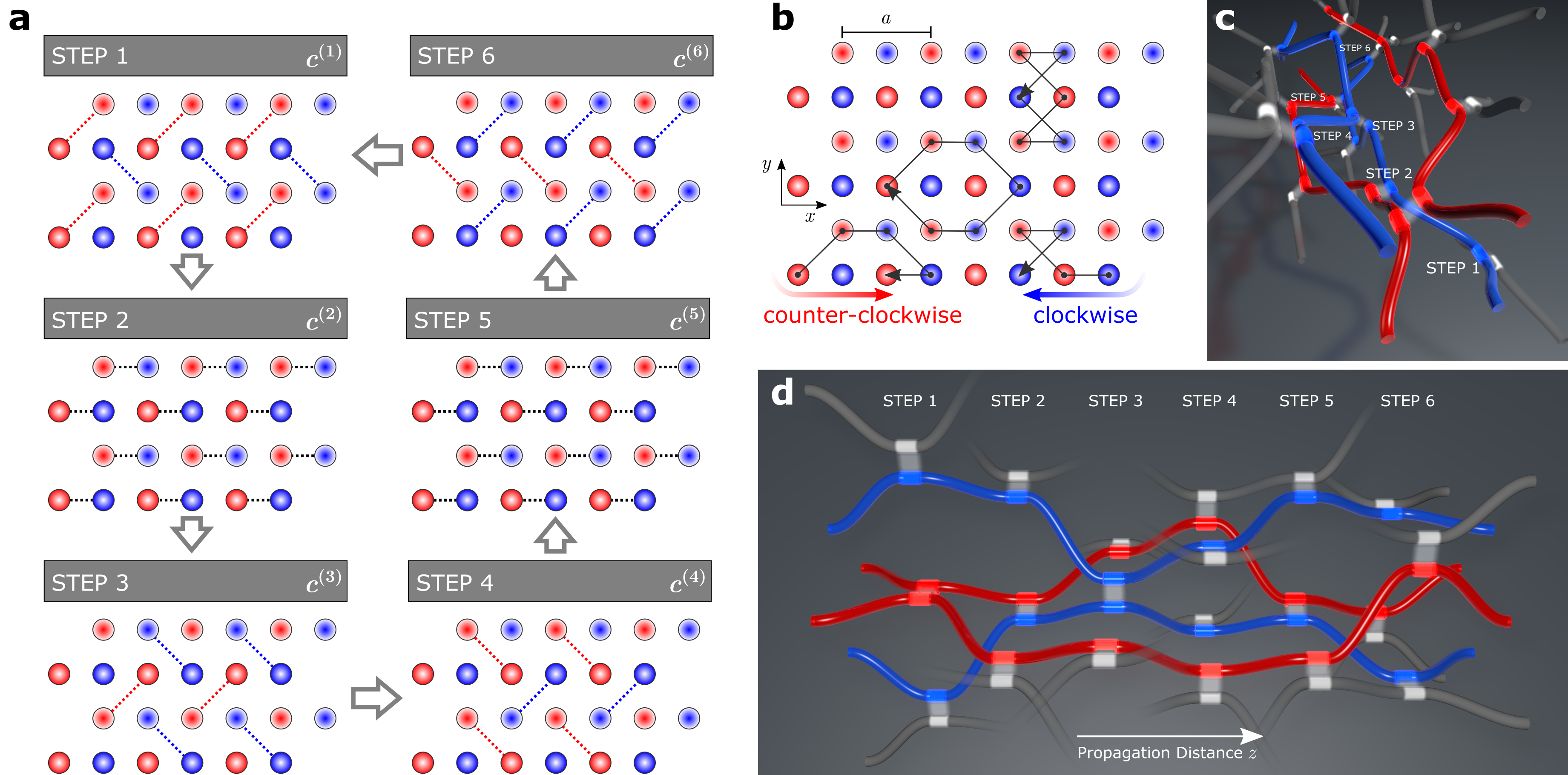}
         \vspace{-8pt}
 	\caption{\label{fig2}\textbf{Schematic of the driving protocol.} (a)~Over the course of one full driving period $T$, adjacent lattice sites are selectively coupled with identical coefficients $c^{(j)}$ in a sequence of six distinct steps $j=1, \dots, 6$. (b)~For full coupling ($c^{(j)}=3\pi/T$), bulk transport is entirely suppressed. The chiral edge states travel along the lattice perimeter in clockwise and counter-clockwise direction, respectively. Blue/red arrows indicate the movement of the edge states. (c),(d)~Schematic of the unit cell geometry (red/blue) embedded within the surrounding waveguide lattice (grey). The hopping region of each coupling section is highlighted by semi-transparent ribbons. The experimental lattice constant is $a=80\,$\textmu m.
}
\end{figure}

For the sake of brevity, the explicit formulation of the associated lattice Hamiltonian $H(t)$ has been relegated to the Supplementary Information. Here, we will instead highlight its fundamental properties: Being of Floquet-type, the Hamiltonian is periodic in time, $H(t+T) = H(t)$, with the driving period $T$. Moreover, $H(t)$ obeys the fermionic TRS relation
\begin{equation}\label{eq:TRS}
 \Theta H(t) \Theta^{-1} = H(T-t) \;,
\end{equation}
where $\Theta$ is an anti-unitary operator with $\Theta^2= -1$. In the (pseudo-) spin interpretation of the two sublattices, we have $\Theta = \sigma_y \, \mathcal K$, with the Pauli matrix $\sigma_y$ and the operator $\mathcal K$ that represents complex conjugation. This symmetry brings about Kramers degeneracy and, in turn, the desired counter-propagating photonic chiral edge states. It should be highlighted that, in contrast to conventional fermionic TIs, this degeneracy is not an intrinsic property of the propagating excitations, but rather associated with the underlying bipartite sublattice structure.

Neither the Chern number $\mathcal C$ \cite{TKNN} nor the Kane-Mele $\mathbb{Z}_2$ invariant $\nu_\mathrm{KM}$ \cite{KaneMelePRL, Fu} are appropriate topological quantities for TIs based on a multi-step driving protocol~\cite{KitagawaPRB}. Instead, the existence of Floquet topological phases is linked to the $\mathcal W$-invariant~\cite{Lindner}. The $\mathcal W$-invariant counts the net topological charge of degeneracy points of the propagator $U(t)= \mathcal T \exp(-\mathrm i\int_0^t H(\tau) \, \mathrm d \tau)$ ($\mathcal T$ denotes time-ordering, and $\hbar$ is set to one) \cite{ Nathan, HockendorfPRB,HockendorfJPA}. However, with TRS, such degeneracies occur in pairs with opposite topological charge. Their contributions therefore cancel and the $\mathcal W$-invariant vanishes, just as the Chern number in a conventional $\mathbb{Z}_2$ insulator. This remains the case even if the Floquet system supports counter-propagating chiral edge states. As it turns out, the existence of non-trivial Floquet topological phases with fermionic TRS is linked to a new $ \mathbb{Z}_2$ invariant $ \nu_\mathrm{TR}$ \cite{ Nathan, HockendorfPRB}, which is connected to the Kane-Mele invariant $\nu_\mathrm{KM}$ in a similar way as the $\mathcal W$-invariant is related to the Chern number. In particular, the $\mathbb{Z}_2$ Floquet TI introduced in this work is characterised by $\mathcal C =0$, $\mathcal W = 0$, $\nu_\mathrm{KM}=0$, but $\nu_\mathrm{TR}=1$. In other words, our system exhibits a non-trivial topological phase with counter-propagating chiral edge states that are protected by TRS and cannot exist in its absence. This phase would be absent without TRS. A detailed overview and comparison of the previously discussed topological invariants is given in the Supplementary Information.

As testbed for the practical implementation and experimental verification of our protocol, we choose an optical platform: lattices of evanescently coupled laser-written waveguides \cite{SzameitJPB}. Light evolves in these structures according to a Schrödinger-type equation, which reads
\begin{equation}
\mathrm i\frac{\mathrm d}{\mathrm dz}\psi_m(z) = \epsilon_m(z)\psi_m(z) + \sum_{k \in \langle m \rangle} c_{k,m}(z)\psi_k(z)
\end{equation}
in the tight-binding approximation.
Here, $\epsilon_m(z)$ is the on-site potential of waveguide $m$, $\psi_m$ represents the field amplitude of its guided mode, $c_{k,m}(z)$ denotes the hopping (coupling) to the nearest neighbour $k$, and the propagation distance $z$ serves as the evolution coordinate. In the summation, $\langle m \rangle$ denotes the nearest neighbours of the $m$th waveguide. Importantly, in the system under consideration, the values of the couplings $c_{k,m}(z)$ differ in each step of the full sequence: As illustrated in Fig.~\ref{fig2}, interactions generally have to be avoided except for the steps that necessitate hopping between two given waveguides. For simplicity, we will refer to the couplings and potentials in step $j$ as $c^{(j)}$ and $\epsilon^{(j)}$.

Implementation of our driving protocol yields the structure depicted in Figs.~\ref{fig2}(c,d): Here, a single unit cell is shown in transverse cross section (c) and longitudinal cross section (d), respectively. For our experiments, we fabricated a lattice spanning four by three unit cells in the $x-y$-plane and three driving cycles along $z$. In real-world units, the unit cell transversely extends over $a^2=80\times80\,$\textmu$\mathrm{ m}^2$ (see Fig.~\ref{fig2}(b)), and $T=4.44\,\mathrm{cm}$ along the propagation direction $z$. Further details of the fabrication, in particular the explicit values of the couplings $c^{(j)}$ and potentials $\epsilon^{(j)}$, are given in the Methods section.

Our samples were characterised by injecting light from a Helium-Neon laser into a single site of the lattice through a $10\times$ microscope objective. We subsequently observed the result of the dynamics after three driving periods, i.e. at the end of the $15\, \mathrm{cm}$ glass sample, by imaging the output facet onto a CCD camera. The recorded intensity patterns were processed to remove excess noise, and normalised to allow for a meaningful quantitative comparison between individual measurements. Figure~\ref{fig3} summarises the results of this experiment. The upper row illustrates the case of a general spin rotation, where the bulk bands are dispersive (Fig.~\ref{fig3}(a)). In the lower row we present the results obtained for a spin flip, where the bands in principle exhibit no dispersion. In both cases, we observe an edge state moving clockwise (in blue) and another one moving counter-clockwise (in red) as shown in Figs.~\ref{fig3}(b,c) and Figs.~\ref{fig3}(e,f). Extended sets of experimental images for both cases are provided in the Supplementary Information.

\begin{figure}
    	\centering
 	\includegraphics[width=1\textwidth]{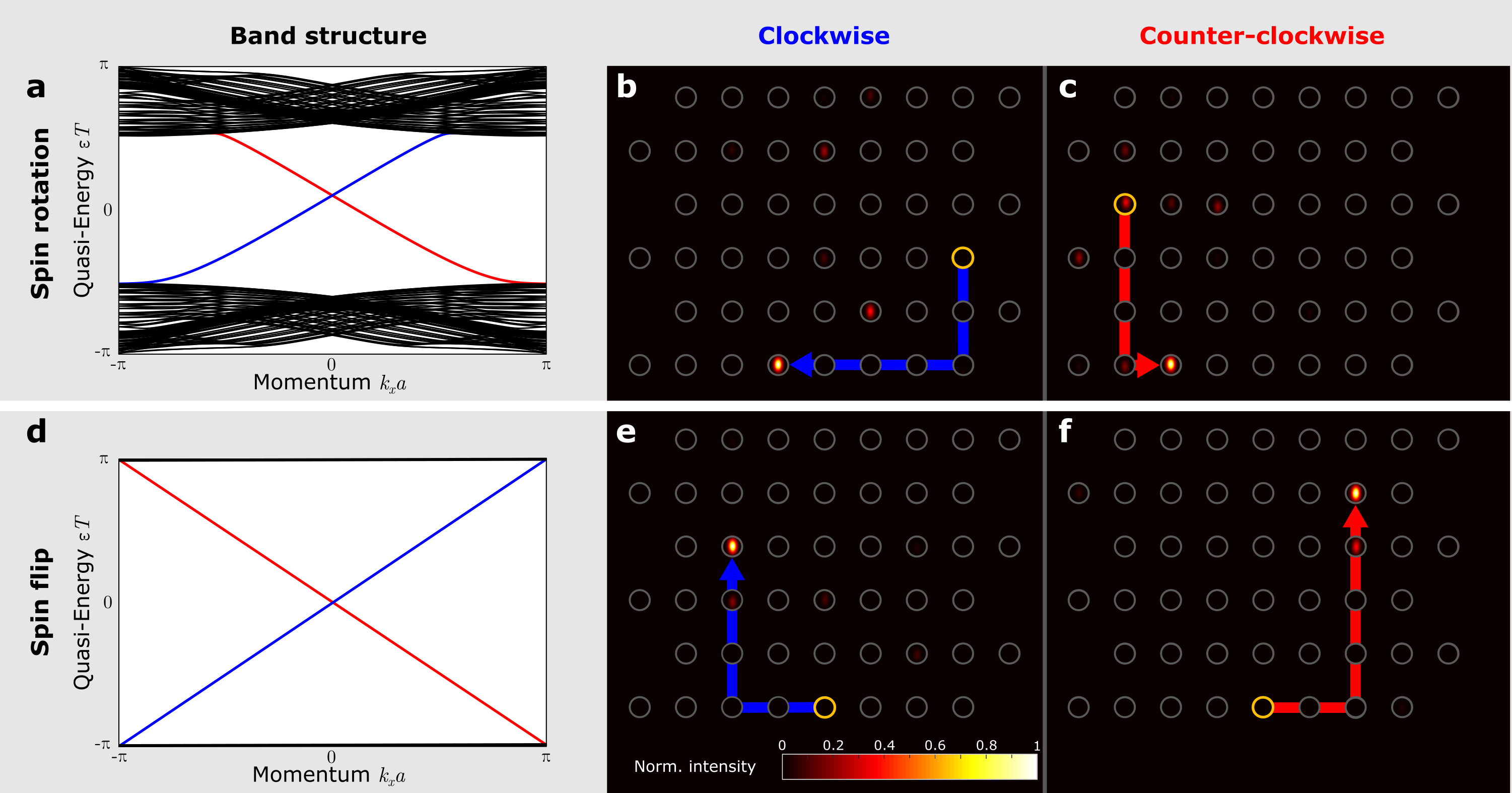}
         \vspace{-8pt}
 	\caption{\label{fig3} \textbf{Experimental demonstration of counter-propagating bosonic edge modes.} (a)~Band structure for the case of a spin rotation. The bulk modes are plotted in grey, whereas the chiral edge modes are shown in blue (clockwise) and red (counter-clockwise), respectively. The associated output intensity distribution after three full driving periods for single-site excitation are shown in (b,c). Grey circles indicate the waveguide positions, whereas the excited sites are marked in yellow. The effective trajectories of the propagating modes are visualised by broad blue/red arrows.
 (d)~Band structure in case of a spin flip. The associated output intensity distribution after three full driving periods for single-site excitation are shown in (e,f). Note that the entirely flat bulk bands go along with chiral edge states with linear dispersion: The single-site excitations no longer spread out along their trajectory, as it was the case for spin rotation (a-c).
}
\end{figure}

Whereas the very existence of two counter-propagating chiral edge states is already a strong indication of Kramers degeneracy, and, by extension, of fermionic TRS, a direct demonstration is within the scope of this experiment. Our line of reasoning relies on the fact forward propagation through the waveguide structure (along the positive z direction) in a system with TRS is intricately linked to backward propagation (along $-z$). It should be emphasised that backward propagation in itself is not identical to time reversal, which cannot be achieved by merely exciting the opposite end of the sample. However, if the system indeed obeys fermionic TRS $(\Theta=\sigma_y\mathcal{K})$, the backward propagator $\tilde U(T)$ is related to the previously defined forward propagator $U(T)$ via the relation $\tilde{U}(T)=\sigma_y U(T)\sigma_y^{-1}$. The mathematical details behind this argument are provided in the Supplementary Information.

Now consider the output states $|\psi_{\textrm{out}}(\phi) \rangle = U(T) |\psi_\mathrm{in}(\phi)\rangle$ and $|\tilde{\psi}_{\textrm{out}}(\phi) \rangle = \tilde U(T)  |\psi_\mathrm{in}(\phi)\rangle$ that evolve from either forward or backward propagation of an input state $|\psi_\mathrm{in}(\phi)\rangle$. In our experiments, a suitable input state $|\psi_\mathrm{in}(\phi)\rangle$ spanning two adjacent (`red' and `blue') waveguides with the same amplitude 158 but a relative phase $\phi$ is synthesised with a spatial light modulator (SLM)
159 as illustrated in Fig.~\ref{fig4}. Note that the same corresponding waveguides are excited in both forward and backward propagation (see Fig.~\ref{fig4}(a)). For both
161 directions, we extract two intensity distributions $\tilde I^\mathrm{R}(\phi)$ or $I^\mathrm{B}(\phi)$ from the observed output states and track their dependence on the relative phase $\phi$ of the input state. The relation between the forward and backward propagator given above then readily translates into
\begin{equation}
\tilde{I}^\mathrm{R}(\phi) =	I^\mathrm{B}(\pi -\phi)\label{eq:fTRS_phaseshift}
\end{equation}
for the output intensities. Notably, this expression is unique to fermionic TRS (see the Supplementary Information for a detailed derivation and discussion). As our experiments do indeed faithfully reproduce the characteristic phase shift $\phi \mapsto \pi - \phi$ as well as the exchange of intensities between the two sublattices predicted by Eq.~\eqref{eq:fTRS_phaseshift} (see Fig.~\ref{fig4}(a)), they  unequivocally confirm the presence of fermionic TRS in our system.

\begin{figure}
    	\centering
 	\includegraphics[width=1\textwidth]{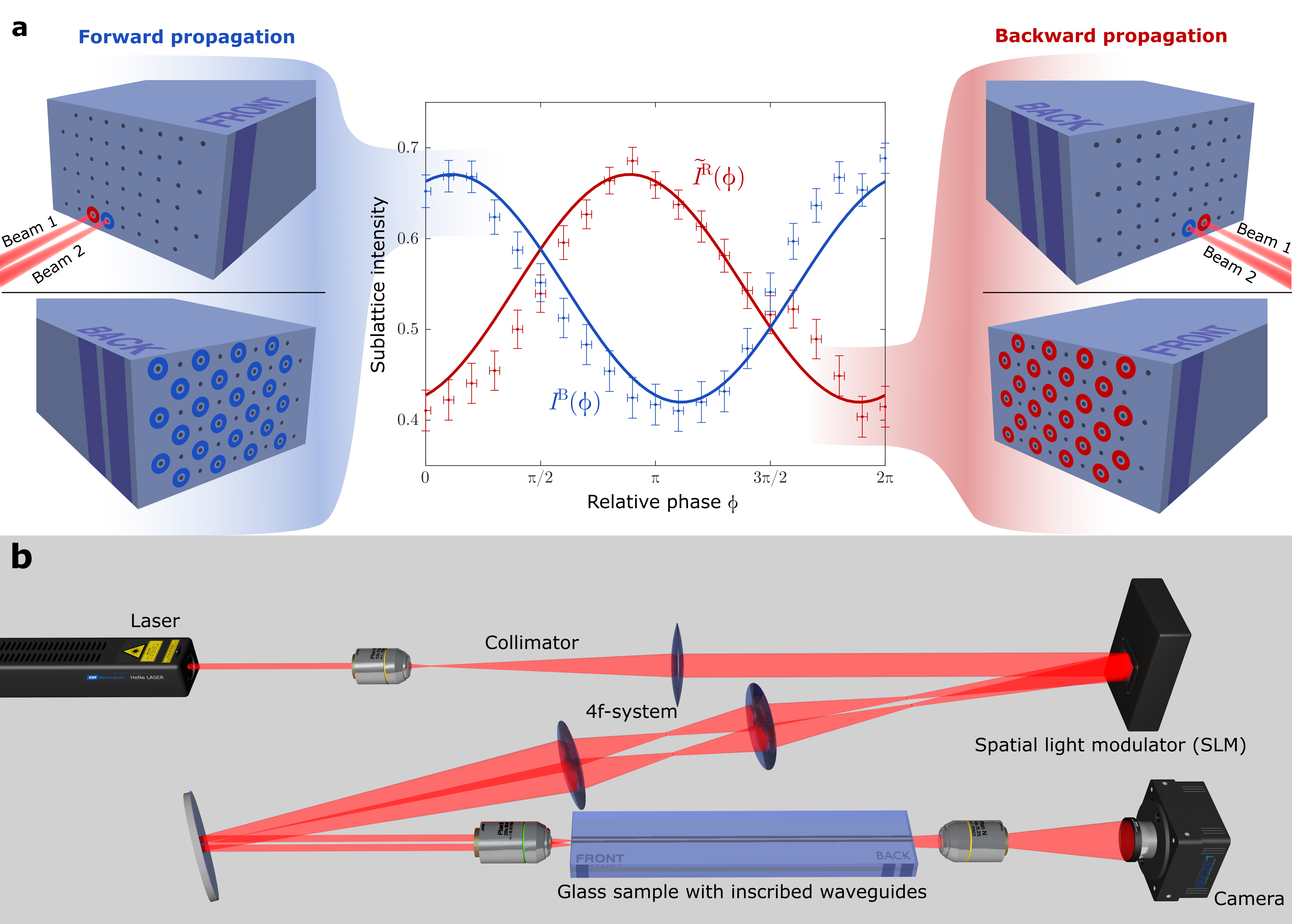}
         \vspace{-8pt}
 	\caption{\label{fig4} \textbf{Experimental verification of fermionic time-reversal symmetry.} (a)~Dependence of the sublattice population on the relative phase of a dual-site excitation (Solid lines: Numerical calculations. Dots: Measured values with error bars). The comparison between forward and backward propagation confirms the predicted characteristic phase shift between the behaviour of the output intensities in sublattices R (red) and B (blue), respectively.
 (b)~Schematic of the experimental setup. The spatial light modulator (SLM) synthesises two phase-shifted beams, which are rescaled and imaged onto the sample, and the resulting output intensity distributions are recorded by a CCD camera.}
\end{figure}

In summary, we have shown that Kramers degeneracy associated with fermionic time-reversal symmetry can be effectively realised for bosonic systems by mapping the spin degree of freedom onto the underlying lattice. The resulting structure is described by a $\mathbb{Z}_2$-type topological invariant and, as such, exhibits two counter-propagating chiral edge states. While we chose an optical platform for this proof of principle, the presented protocol is general and can be readily adopted in any bosonic wave system. In this vein, we expect the experimental realization of a photonic system with fermionic TRS to stimulate fruitful theoretical and experimental efforts to illuminate the role of $\mathbb{Z}_2$-type invariants in bosonic topological systems in greater detail. Fascinating topics waiting to be explored include the impact of interactions in optical, atomic and condensed-matter systems on topological phases with TRS, the possibility of similar phases persisting in the quantum many-body regime, and the potential interplay with non-Hermiticity. The answers to these question, and many more, are now within the reach of experiments.

\subsection*{Acknowledgements}
AS gratefully acknowledge financial support from the Deutsche Forschungsgemeinschaft (grants SZ 276/9-1, SZ 276/19-1, SZ 276/20-1) and the Alfried Krupp von Bohlen und Halbach Foundation. The authors would also like to thank C. Otto for preparing the high-quality fused silica samples used in all experiments presented here.

\subsection*{Author contributions}
The theoretical concept was proposed by BH, AA, and HF.
LM designed the photonic implementation and carried out the experiments with MK and TB.
LM evaluated the measurements and prepared the figures with MH.
The theory was developed by BH and AA.
HF and AS directed the efforts of their respective groups.
The manuscript was primarily written by LM, BH, MH and AA. All authors discussed the results and finalised the manuscript.

%

\subsection*{Competing interests}
The authors declare no competing interests.

\subsection*{Methods}
\subsubsection*{Sample fabrication}
The waveguide lattices used in our experiments were fabricated by means of the femtosecond laser direct writing technique \cite{SzameitJPB}. Pulses from a Ti:Sapphire amplifier system (\textsc{Coherent} Mira 900/RegA 9000, wavelength 800 nm, repetition rate 100 kHz, pulse energy 450 nJ) are focused into the bulk of a fused silica wafer (\textsc{Corning} 7980, dimensions $1\times20\times150$ mm$^3$) by means of a $20\times$ microscopy objective ($0.35\,$NA). A three-axis positioning system (\textsc{Aerotech} ALS 130) was used to inscribe extended lines of permanent refractive index modifications on the order of $7\cdot 10^{-4}$ by translating the sample with respect to the focal spot. At the probe wavelength of 633 nm, these waveguides exhibit a mode field diameter of $10.4\,\,$\textmu m $ \times \ 8\,$\textmu m and anisotropic coupling in the $x-y$-plane. The discrete hopping steps were implemented via dedicated directional couplers (length $6\,\mathrm{mm}$) connected by sinusoidal fan-in/fan-out branches mediating the transitions (length $1.4\,\mathrm{mm}$) of subsequent steps.
Moreover, we made use of the fact that the trajectories of these transition sections can readily be fashioned with precisely defined differences in their overall optical path lengths, which in turn allows propagating light to accumulate the same additional phases that a detuned coupler would produce. In this vein, we are able to selectively implement diagonal terms in the discrete Hamiltonian without having to physically change the on-site potential.
The spin flip case was achieved with a coupling separation of $11.6\,$\textmu m (diagonal interactions, $c^{(1,3,4,6)}=3\pi/T$) and  $10.9\,$\textmu m (horizontal interactions, $c^{(2,5)}=3\pi/T$). The spin rotation case was in turn implemented with separations of  $11.6\,$\textmu m, $12.5\,$\textmu m and $10.2\,$\textmu m for $c^{(1,3,4,6)}=5\pi/2T$, $c^{(2)}=2\pi/T$, $c^{(5)}=4\pi/T$ respectively and an effective on-site potential $\epsilon^{(1,3,4,6)}=3/2T$. The lattice for the probing of the TRS was manufactured with only one driving period and the parameters $c^{(1,3,4,6)}=9\pi/4T$, $c^{(2)}=2\pi/T$, $c^{(5)}=4\pi/T$ and $\epsilon^{(1,3,4,6)}=4/T$. The suppression of undesirable interactions was ensured by increasing the waveguide separation to $40\,$\textmu m in the inert regions.

\subsubsection*{Probing the lattice dynamics}
The samples were illuminated by 633\,nm light from a Helium-Neon laser (\textsc{Melles-Griot}, $35$\,mW). For the demonstration of the counter-propagating modes, a single lattice site was excited with a $10\times$ microscope objective ($0.25$\,NA). Another $10\times$ microscope objective was used to image the output facet onto a CCD camera (\textsc{Basler} Aviator). The recorded images were post-processed to reduce noise and filtered to extract the actual modal intensities while reducing the influence of background light.

The two-site excitations for the verification of TRS were synthesised by means of a spatial light modulator (\textsc{Hamamatsu} LCOS-SLM X0468-02) with a holographic pattern comprising two separated Fresnel lenses. Additionally, these patterns were offset to impart a relative phase onto these two beams. A 4f-setup (focal lengths 1000$\,$mm and 125$\,$mm) and a $20\times$ microscope objective (NA$=0.40$) served to scale down the beam diameters and separation to excite two adjacent waveguides. The resulting output intensity distributions were similarly recorded and post-processed to extract the data plotted in Fig.\ref{fig4}(a). Note that in order to obtain a non-zero contrast from the sine/cosine shaped intensity-phase-dependences, coupling steps 1, 3, 4 and 6 necessarily require non-zero diagonal entries in the Hamiltonian. In line with the approach described above, these were implemented via geometric path differences of $9.6\,$\textmu m (transitions from step $1\rightarrow2$ and $2\rightarrow3$) and $9.9\,$\textmu m ($4\rightarrow5$ and $5\rightarrow6$), which would in the conventional realization correspond to a detuning of $4/T$ within the couplers of steps 1, 2, 3 and 4.

\subsubsection*{Numerical calculations}

The band structures in Figs.~\ref{fig1b},~\ref{fig3} were obtained by diagonalizing the Floquet-
255 Bloch propagator $U(\mathbf k,T)$ after one driving period $T$,
which provides the quasi-energies $\varepsilon$ as a function of momentum $k_x, k_y$.
The propagator $U(\mathbf k,T)$ was computed numerically with the Bloch Hamiltonian $H(\mathbf k, t)$ of the driving protocol in momentum space (the explicit expression for $H(\mathbf k, t)$ is given in the Supplementary Information).
To compute the dispersion of the edges states in Figs.~\ref{fig1b},~\ref{fig3}, the Floquet propagator on a semi-infinite ribbon was computed as a function of momentum $k_x$ or $k_y$ parallel to the edges. The width of the ribbon was chosen as $15$ unit cells,
and only the edge states on one edge of the ribbon were included in the figures.
Further details on the ribbon geometry are provided in the Supplementary Information.

For the numerical results in Fig.~\ref{fig4}(a) (solid curves for $I^{\mathrm{B}}(\phi), \tilde I^{\mathrm R}(\phi)$) the Floquet propagators $U(T)$ of forward and $\tilde U(T)$ of backward propagation were computed with the lattice Hamiltonian $H(t)$ of the driving protocol on a finite lattice with $4\times 3$ unit cells in the $x$-$y$-plane, as in Figs.~\ref{fig2},~\ref{fig3}.
%
%

In all computations, the parameters $c^{(j)}$, $\epsilon^{(j)}$ of the driving protocol have been set to the relevant experimental values specified previously for the spin rotation (for Figs.~\ref{fig1b},~\ref{fig3}(a)) and spin flip (for Fig.~\ref{fig3}(d)) case, or for probing of TRS (for Fig.~\ref{fig4}(a)).

\newpage
	\cleardoublepage
	\setcounter{figure}{0}

	\renewcommand{\figurename}{Figure}
	
	
	\begin{center}
		\text{\normalsize Supplementary Information for}\\
		\vspace{3mm}
		\text{\large Fermionic time-reversal symmetry in a photonic topological insulator}
		\vspace{4mm}
		
\author{Lukas J. Maczewsky$^{1,\ast}$, Bastian Höckendorf$^{2,\ast}$,\\ Mark Kremer$^{1}$, Tobias Biesenthal$^{1}$, Matthias Heinrich$^{1}$\\ Andreas Alvermann$^{2}$, Holger Fehske$^{2}$, and Alexander Szameit$^{1}$\\
\normalsize{$^1$ Institut f\"ur Physik, Universit\"at Rostock, Albert-Einstein-Str. 23,}\\ \normalsize{18059 Rostock, Germany.}\\
\normalsize{$^2$ Institut für Physik, Universität Greifswald, Felix-Hausdorff-Str. 6,}\\ \normalsize{17489 Greifswald, Germany.}\\
\normalsize{$^\ast$ These authors contributed equally.}\\
\textit{\normalsize{alvermann@physik.uni-greifswald.de; alexander.szameit@uni-rostock.de}}
\\
}
		
		\vspace{5mm}
	\end{center}
	
	\renewcommand{\figurename}{Figure}

\setcounter{equation}{0}
\setcounter{section}{0}
\setcounter{figure}{0}
\setcounter{table}{0}
\setcounter{page}{1}
\makeatletter

\renewcommand{\thefigure}{S\arabic{figure}}
\renewcommand{\thetable}{S\arabic{table}}

\setcounter{enumi}{1}
\renewcommand{\theequation}{S\Roman{enumi}.\arabic{equation}}
\renewcommand{\thesection}{\Roman{section}}

\section{Experimental techniques}

\subsection{Implementation of on-site potentials}
While the femtosecond laser inscription technique is capable of directly and precisely modulating the effective index of the fabricated waveguides via the exposure parameters (pulse energy, writing velocity), we followed a different approach in this work to selectively implement on-diagonal terms in the discrete Hamiltonian. Instead of writing detuned couplers, i.\,e. evanescently interacting waveguides with different effective refractive indices, we designed the trajectories of the transition sections between subsequent steps such that precisely defined differences in their overall optical path lengths allow propagating light to accumulate the same additional phases that physically detuned couplers would produce. This technique is of particular importance for the verification of time reversal symmetry, since
in order to obtain a non-zero contrast of the sine/cosine shaped intensity-phase-dependences, coupling steps 1, 3, 4 and 6 necessarily require detuned on-diagonal entries of the Hamiltonian. In line with the approach described above, these were implemented via geometric path differences of $9.6\,$\textmu m (transitions from step $1\rightarrow2$ and $2\rightarrow3$) and $9.9\,$\textmu m ($4\rightarrow5$ and $5\rightarrow6$), which would in the conventional realization correspond to a detuning of $4/T$ within the couplers of steps 1, 2, 3 and 4 (see Fig. \ref{figsup3}).

\begin{figure}
    	\centering
 	\includegraphics[width=0.68\textwidth]{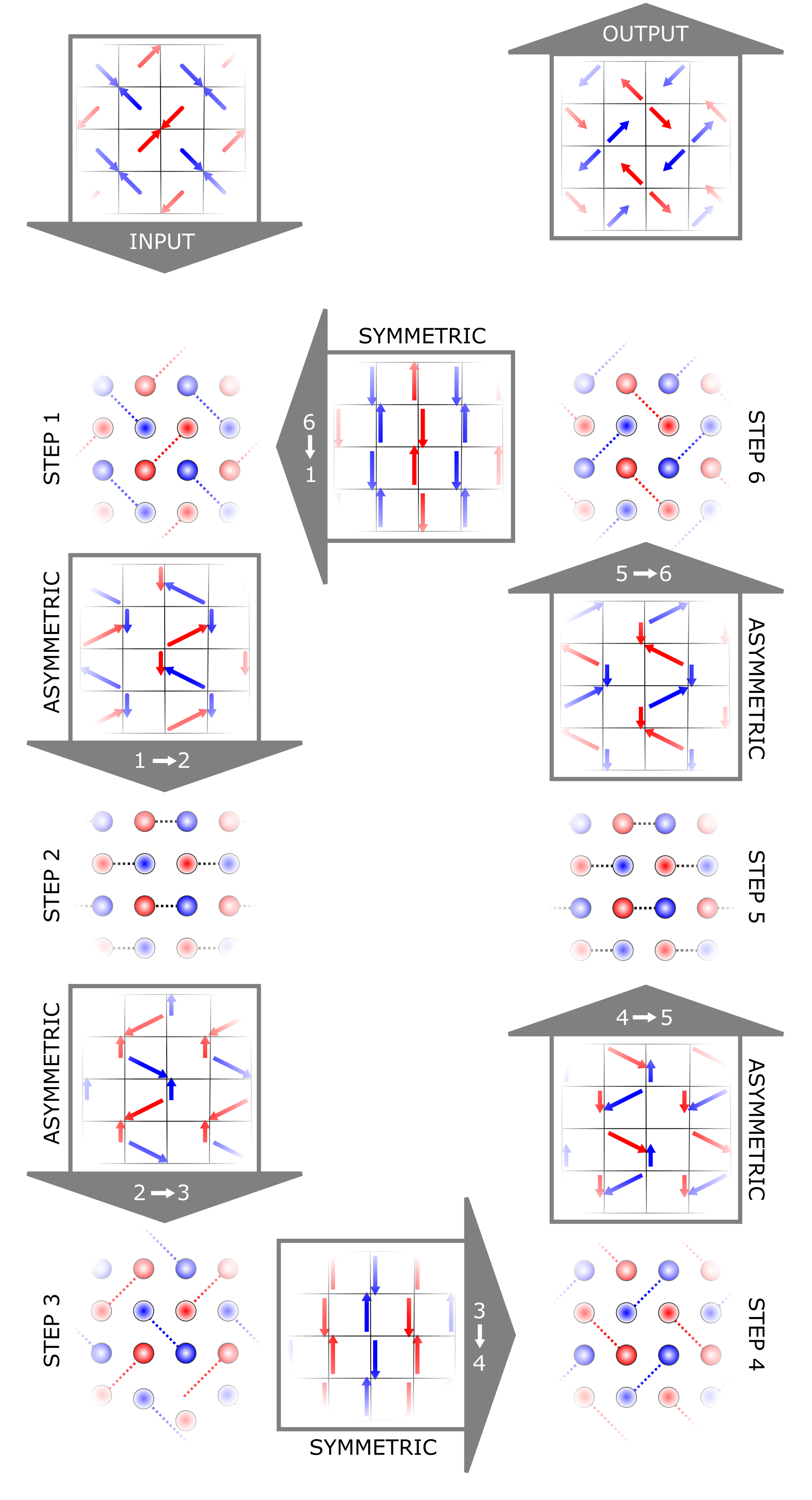}
         \vspace{-8pt}
 	\caption{\label{figsup3} \textbf{Implementation of the on-site potential.} The discrete driving protocol of Fig.~2(a) of the main text is combined with trajectories of the waveguides between the hopping steps. The trajectories are marked by the red and blue arrows. The length of these arrows corresponds to the optical path length of the light guided by the waveguides. The asymmetric path lengths are clearly visible in the transitions from step $1\rightarrow2$ and $2\rightarrow3$.}
\end{figure}

\subsection{Additional edge state measurements}
As further evidence for the predicted edge state behaviour in our system, Fig.~\ref{figsup1} and Fig.~\ref{figsup2} show the output intensity profiles for additional single-site excitations beyond the ones shown in Fig.~4. Note that the spin flip case (Fig.~\ref{figsup1}) is characterized both by chiral edge transport (panels a/e,~b/f), as well as a flat bulk band (Fig.~3(d)). The latter is responsible for the localized bulk excitations (panels c/g and d/h). The more general spin-rotation case (Fig.~\ref{figsup2}) continues to support the edge states. However, owing to the non-zero curvature of their trajectories through the band diagram (Fig.~4(a)), these edge states exhibit non-uniform transverse velocities. As a result, single-site edge excitations remain decoupled from the bulk, but are subject to a certain degree of dispersive broadening as they propagate along the edges.

\begin{figure}
    	\centering
 	\includegraphics[width=0.7\textwidth]{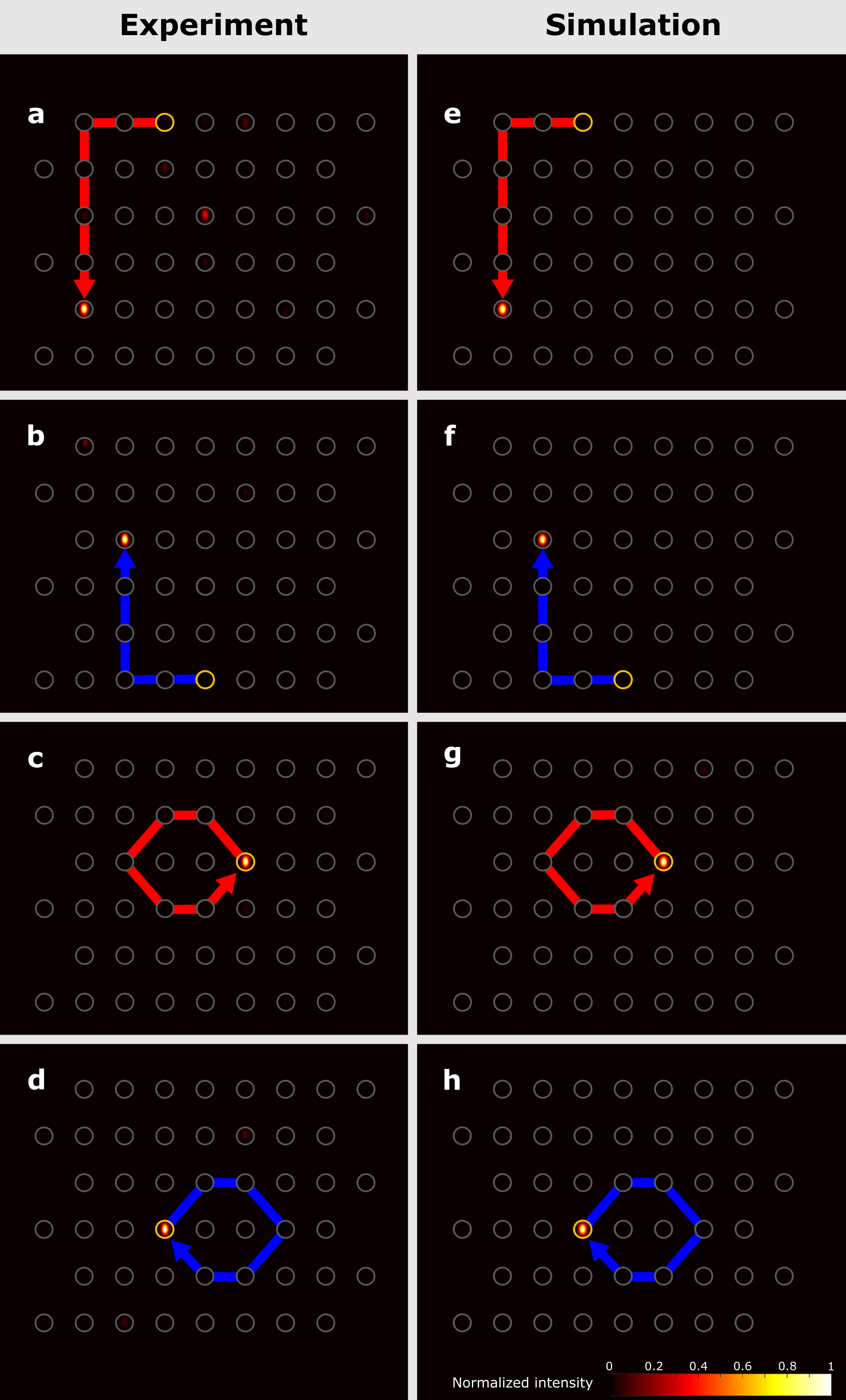}
         \vspace{-8pt}
 	\caption{\label{figsup1} \textbf{Additional data for the spin-flip case.} Shown are the output intensity distributions resulting from excitations of the orange-outlined lattice sites after three full driving periods. The effective wave packet trajectories are indicated by blue and red arrows for clockwise and counter-clockwise propagation, respectively. (a,b) Edge excitations exhibit chiral transport, (c,d) bulk excitations remain effectively localized after each driving period. (e-h)~Corresponding numerical simulations.}
\end{figure}
\begin{figure}
    	\centering
 	\includegraphics[width=0.7\textwidth]{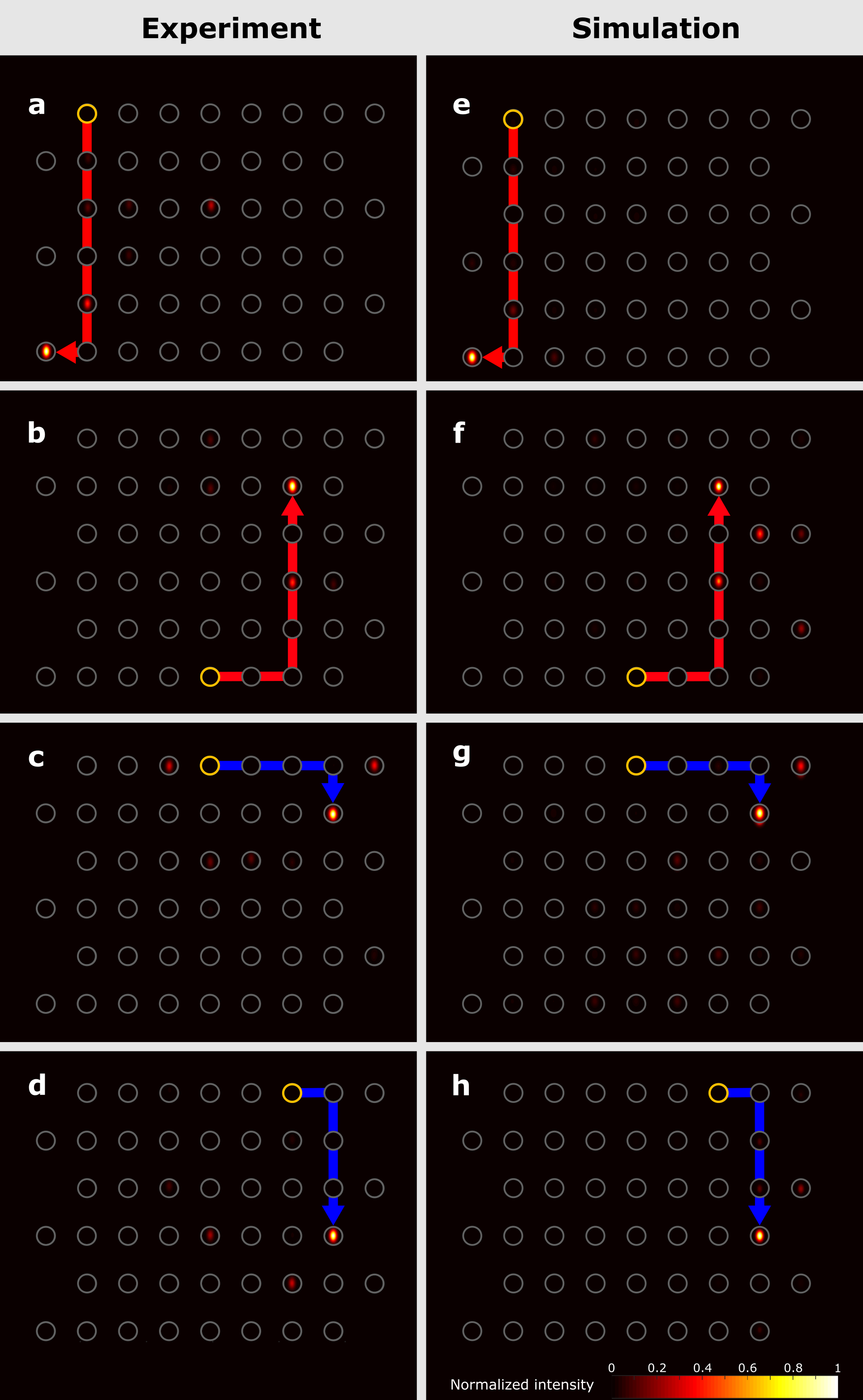}
         \vspace{-8pt}
 	\caption{\label{figsup2} \textbf{Additional data for the spin-rotation case.} Shown are the output intensity distributions resulting from edge excitations of the orange-outlined lattice sites after three full driving periods. The effective wave packet trajectories are indicated by blue and red arrows for clockwise and counter-clockwise propagation, respectively.
The edge states associated with both sublattices R (panels a,b) and B (panels c,d) now exhibit non-uniform transverse velocities, as indicated by a certain amount of wave packet broadening.
(e-h)~Corresponding numerical simulations.}
\end{figure}

\pagebreak
\section{Theory}

\subsection{Construction of the driving protocol}

Our construction of a driving protocol with fermionic time-reversal symmetry (TRS) follows the conceptual idea depicted in Fig.~\ref{theory:fig:concept}.
The driving protocol is based on the square lattice model proposed in Ref.~\cite{Lindner},
which combines the four elementary coupling patterns between adjacent lattice sites
defined in Fig.~\ref{theory:fig:pattern}.
To denote these patterns in the real-space Hamiltonian $H(t)$ of the driving protocol, we use the shorthand graphical
notation
\begin{equation}
\vcenter{\hbox{\includegraphics[width=0.7cm]{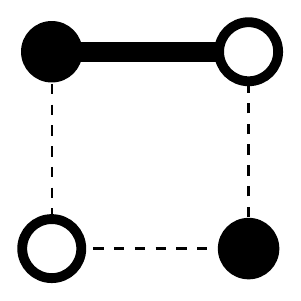}}} \;,
\quad
\vcenter{\hbox{\includegraphics[width=0.7cm]{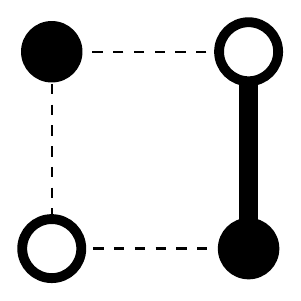}}} \;,
\quad
\vcenter{\hbox{\includegraphics[width=0.7cm]{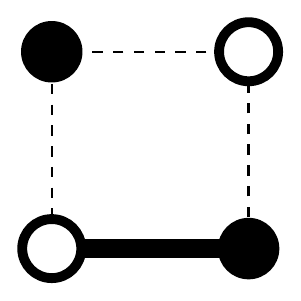}}} \;,
\quad
\vcenter{\hbox{\includegraphics[width=0.7cm]{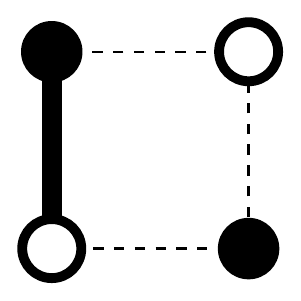}}}
\end{equation}
introduced in this figure. Similarly, we write
\begin{equation}
\vcenter{\hbox{\includegraphics[width=1.0472cm]{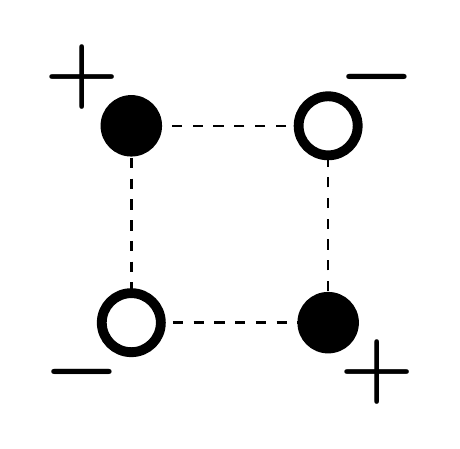}}}
 = \sum\limits_{k,l} \; (-1)^{k+l} | k , l\rangle \langle k,l|
\end{equation}
for a term with alternating on-site potentials.
In this notation, the ket vector $|k,l\rangle$, for $k,l \in \mathbb Z$, denotes the state at the $k$th and $l$th lattice site in horizontal and vertical direction, respectively.
Lattice sites with even $k+l$ are identified with filled circles,
sites with  odd $k+l$ with hollow circles.

\begin{figure}
\hspace*{\fill}%
\includegraphics[width=\linewidth]{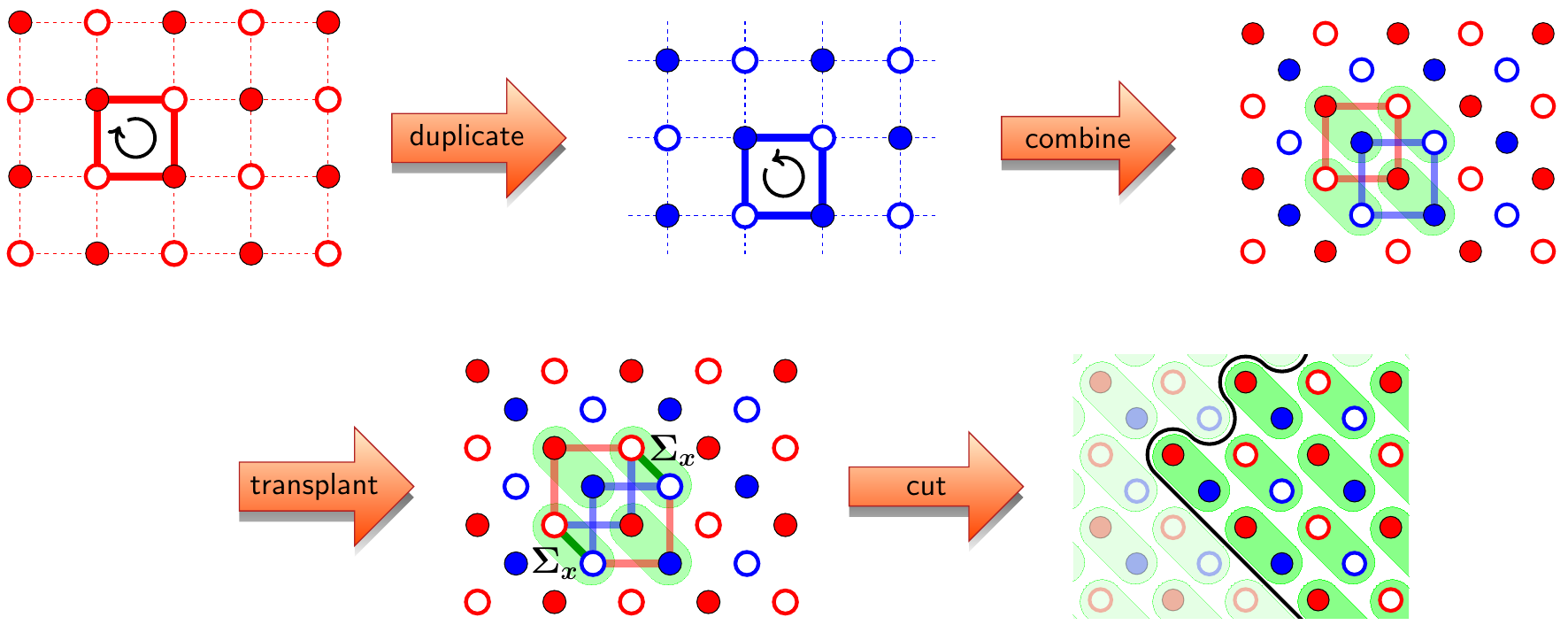}%
\hspace*{\fill}%
\caption{Construction of the TR symmetric driving protocol:
Two copies (``red'' and ``blue'') of a driving protocol with opposite chirality are combined into a centered square lattice. The red/blue sublattice structure can be associated with a pseudo-spin $\tfrac12$, where two neighboring lattice sites are paired (``green'' oval). After rotation by $45^\circ$, this construction gives the protocol depicted in Fig.~3 in the main text.
}
\label{theory:fig:concept}
\end{figure}

\begin{figure}
\hspace*{\fill}%
\includegraphics[width=0.7\linewidth]{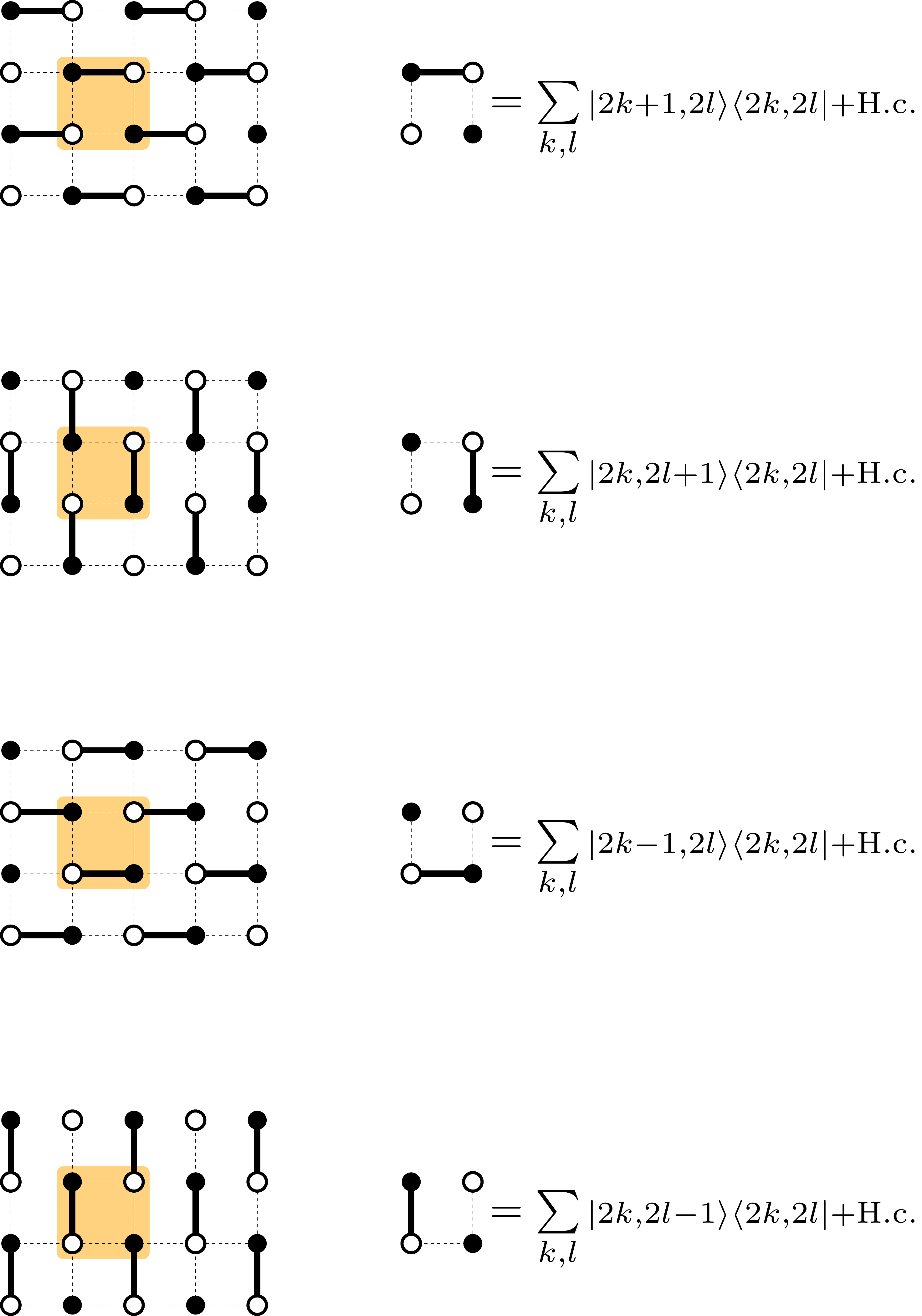}%
\hspace*{\fill}%
\caption{Shorthand graphical notation for the four elementary coupling patterns on the square lattice.
}
\label{theory:fig:pattern}
\end{figure}

\clearpage

If the four coupling patterns are arranged in a periodic sequence, as in the model from Ref.~\cite{Lindner}, the resulting driving protocol implements a Floquet topological insulator with chiral edges states, but non-trivial symmetries cannot be enforced without modification of the protocol~\cite{Carpentier}.

Therefore, to construct a TR symmetric driving protocol, we
\emph{duplicate} the previous non-symmetric model and \emph{combine} the two copies,
as shown in Fig.~\ref{theory:fig:concept}.
 One copy is the mirror image of the other, such that they implement opposite chirality for states on equivalent lattice sites. For the theoretical analysis, it is convenient to associate the two copies with a pseudo-spin $\tfrac 12$,
 where we identify the ``red'' and ``blue'' sublattice of the centered square lattice in Fig.~\ref{theory:fig:concept} with the ``up'' spin state $|{\uparrow}\rangle$ and ``down'' spin state $|{\downarrow}\rangle$, respectively.
In this way, the coupling patterns become associated with the two spin directions.
We have, for example,
\begin{align}
 \vcenter{\hbox{\includegraphics[width=0.7cm]{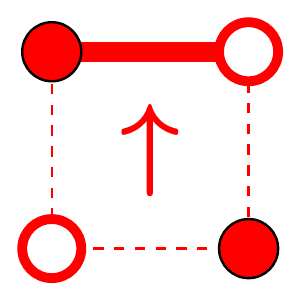}}}
 &= \sum\limits_{k,l} | 2k+1,  2l \rangle \langle 2k, 2l |\otimes  |{\uparrow}\rangle \langle{\uparrow}| + \mathrm{H.c.} \;, \\
\vcenter{\hbox{\includegraphics[width=0.7cm]{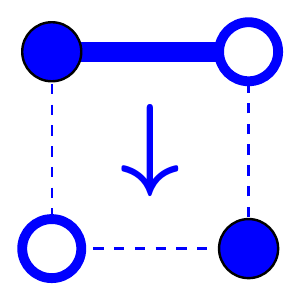}}}
&= \sum\limits_{k,l} | 2k+1,  2l \rangle \langle 2k, 2l |\otimes  |{\downarrow}\rangle \langle{\downarrow}| + \mathrm{H.c.} \;,
\end{align}
and similarly
\begin{equation}
\vcenter{\hbox{\includegraphics[width=1.0472cm]{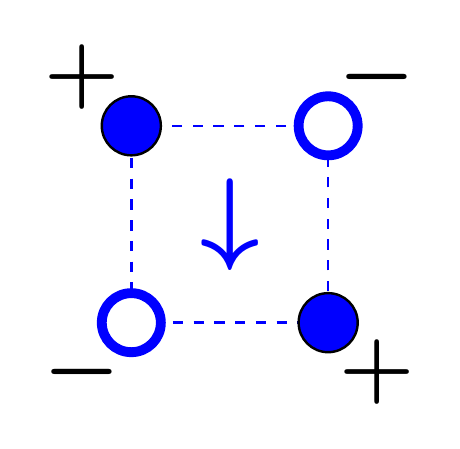}}}
= \sum\limits_{k,l} (-1)^{k+l} | k,  l \rangle \langle k, l |\otimes  |{\downarrow}\rangle \langle{\downarrow}| + \mathrm{H.c.}
\end{equation}
for the potential terms.
These terms preserve the pseudo-spin direction, as expressed by the projections $|{\uparrow}\rangle\langle{\uparrow}| = \tfrac12 (1 + \sigma_z)$ and $|{\downarrow}\rangle\langle{\downarrow}| = \tfrac12 (1 - \sigma_z)$.

To connect the two pseudo-spin directions, or sublattices, steps with a pseudo-spin transformation
\begin{equation}
\sigma_x =  |{\uparrow}\rangle\langle{\downarrow}| + |{\downarrow}\rangle\langle{\uparrow}| \;,
\end{equation}
 need to be included in the driving protocol.
In order to preserve TRS, these steps have to appear pairwise in symmetric position, in our case as steps 2 and 5 of the protocol.

\begin{table}
\caption{Hamiltonian $H(t)$ of the driving protocol in pseudo-spin representation,
using the graphical notation from Fig.~\ref{theory:fig:pattern}.}
\vspace*{1ex}
\hspace*{\fill}%
\includegraphics[width=0.95\linewidth]{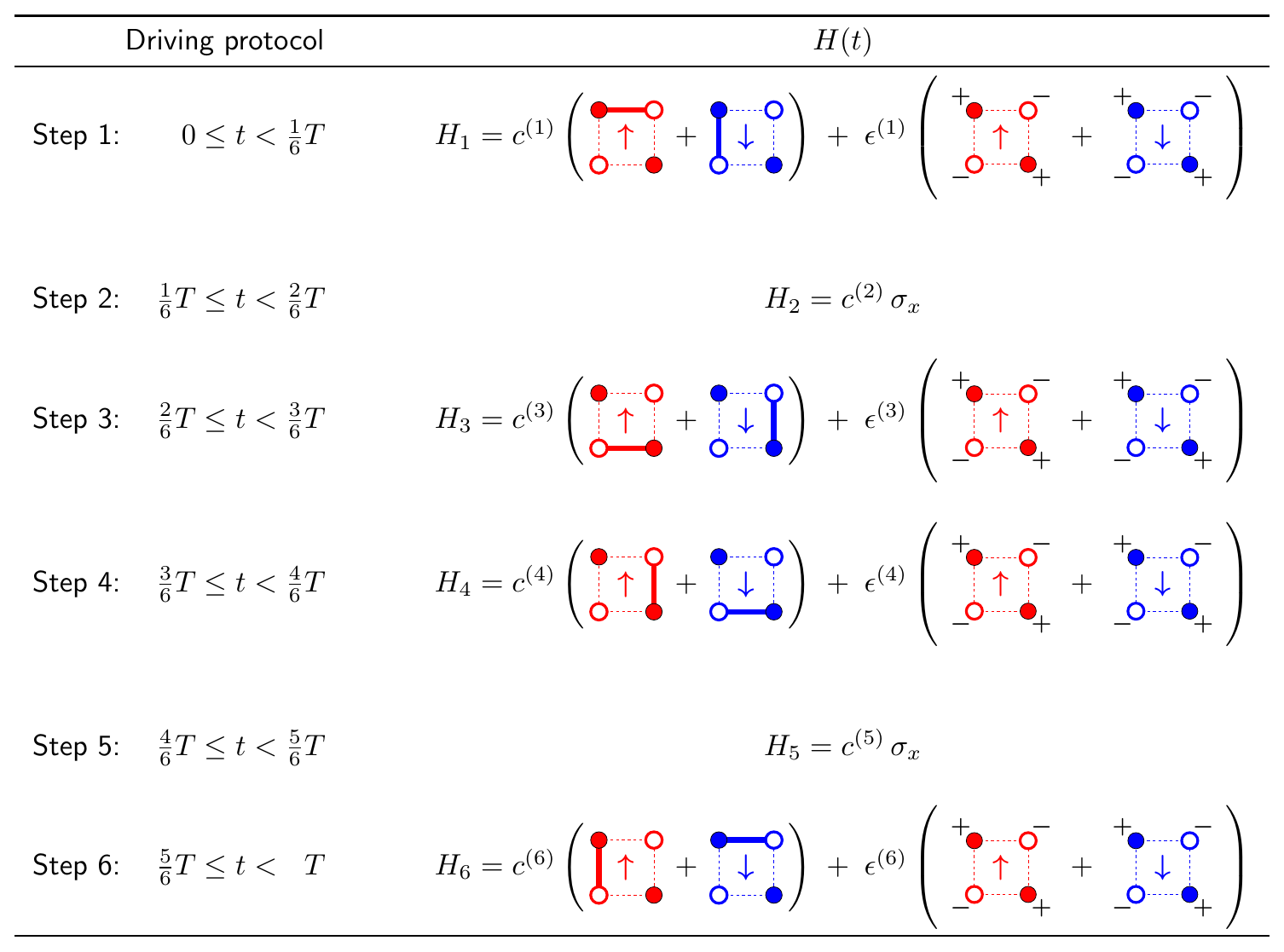}%
\hspace*{\fill}%
\label{theory:tab:Ham}
\end{table}

\begin{table}
\caption{Same as Tab.~\ref{theory:tab:Ham}, now for the Bloch Hamiltonian $H(\mathbf k,t)$ of the driving protocol.}
\vspace*{1ex}
\hspace*{\fill}%
\includegraphics[width=0.95\linewidth]{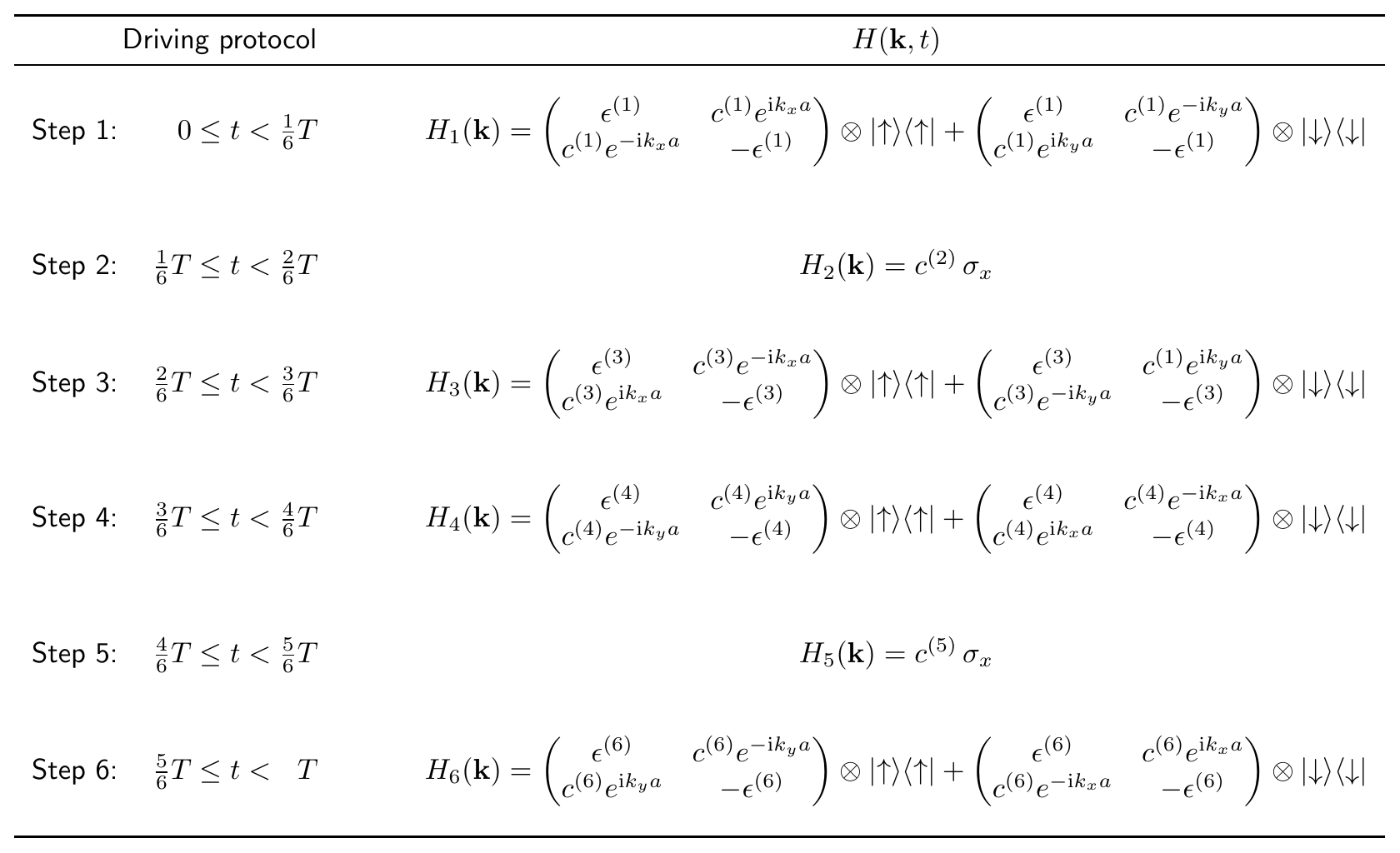}%
\hspace*{\fill}%
\label{theory:tab:Bloch_Ham}
\end{table}

The entire construction results in the driving protocol specified by the time-dependent Hamiltonian
\begin{align} \label{theory:eq:realHamiltonian}
H(t)= H_j \;, \quad \text{ for } \left(n + \tfrac{j-1}{6} \right) T \leq t < \left(n + \tfrac{j}{6}\right) T  \;  \text{ with } \; \ n \in \mathbb N \;,
\end{align}
where the Hamiltonians $H_j$ of each step $j \in\{1,\ldots,6\}$ are listed in Tab.~\ref{theory:tab:Ham}.
By construction, the Hamiltonian is Hermitian and periodic, $H(t+T) = H(t)$. Each period consists of six steps of equal duration $T/6$.
Steps 1, 3, 4 and 6 leave the pseudo-spin unchanged, while steps 2, 5 involve a pseudo-spin rotation. To allow for breaking of particle-hole symmetry, steps 1, 3, 4 and 6 contain additional on-site potentials.
In summary, the driving protocol has ten parameters: six couplings $c^{(j)}$, for $j \in\{1,\ldots,6\}$, and four on-site potentials $\epsilon^{(j)}$, for $j \in\{1,3,4,6\}$.
All parameters, hence also the entire Hamiltonian, are real-valued.
The Hamiltonian in Eq.~\eqref{theory:eq:realHamiltonian} has been used in all numerical calculations
presented in this work, and is the basis of the experimental implementation.

From this Hamiltonian, the Floquet propagator
\begin{equation}\label{theory:floquet_prop}
U(T) = U_6 \, U_5 \, U_4 \, U_3 \, U_2 \, U_1
\end{equation}
is obtained, where the six propagators for each step are defined by  $U_j = \exp \big(-\mathrm i  H_j T/6 \big)$.
For full coupling ($c^{(j)} = \pm 3 \pi /T$, $\epsilon^{(j)}=0$)
the Floquet propagator in the bulk is trivial ($U(T) = \pm \mathbbm 1$).
Especially, steps 2 and 5 correspond to a spin flip $U_{2,5} = \pm \mathrm i \sigma_x$ and thus \emph{transplant} states from one to the other pseudo-spin direction (see Fig.~\ref{theory:fig:concept}).
The introduction of edges gives rise to pairs of edge states with opposite chirality, which move along the trajectories depicted in Fig.~2 in the main text.
Note that an edge must result from a \emph{cut} that preserves TRS,
and does not separate lattice sites that are paired in the pseudo-spin (or red and blue sublattice) representation (see last panel in Fig.~\ref{theory:fig:concept}).

From the real-space Hamiltonian $H(t)$, one obtains the Bloch-Hamiltonian $H(\mathbf k, t)$ in momentum space given in Table~\ref{theory:tab:Bloch_Ham}.
With this Hamiltonian, computation of the bulk band structures in Fig.~2 and Fig.~\ref{theory:fig:bands} (below) is straightforward.

\paragraph{Pseudo-spin to lattice mapping}

As mentioned before, we map the up spin state $|{\uparrow}\rangle$ onto the ``red'' and the down spin state $|{\downarrow}\rangle$ onto the ``blue'' sublattice  to obtain a pure lattice model without pseudo-spin degrees of freedom, which is suitable for a photonic waveguide implementation.
Now, the ket vector $|k,l,R/B\rangle$ carries the sublattice information $R/B$ in addition to the lattice site position $k, l$, and the coupling and potential terms read,
\begin{align}
 \vcenter{\hbox{\includegraphics[width=0.7cm]{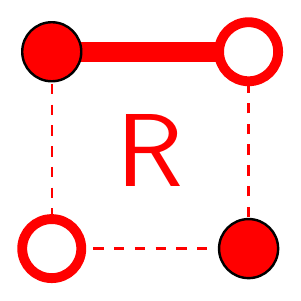}}}
 & \simeq \sum\limits_{k,l} | 2k+1,  2l, R \rangle \langle 2k, 2l, R | + \mathrm{H.c.} \;, \\
\vcenter{\hbox{\includegraphics[width=0.7cm]{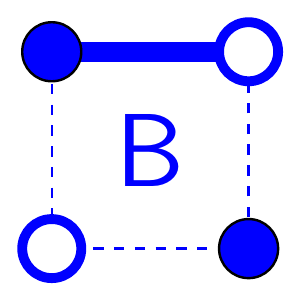}}}
& \simeq \sum\limits_{k,l} | 2k+1,  2l, B \rangle \langle 2k, 2l, B | + \mathrm{H.c.} \;, \\
\end{align}
or
\begin{equation}
\vcenter{\hbox{\includegraphics[width=1.0472cm]{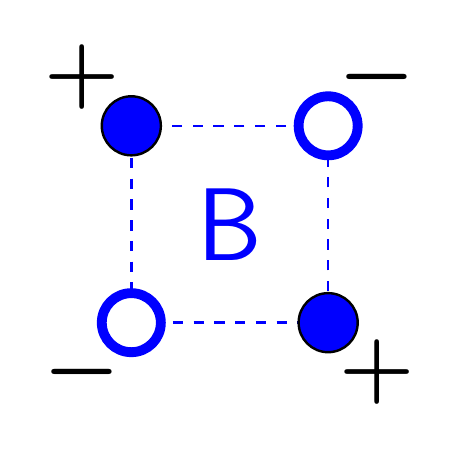}}}
\simeq \sum\limits_{k,l} (-1)^{k+l} | k,  l, B \rangle \langle k, l, B | + \mathrm{H.c.} \;,
\end{equation}
 and similarly for the remaining terms.
The pseudo-spin transformation $\sigma_x$ in steps $2$ and $5$ is replaced by the operator
\begin{align}\label{theory:SigmaX}
\vcenter{\hbox{\includegraphics[width=1cm]{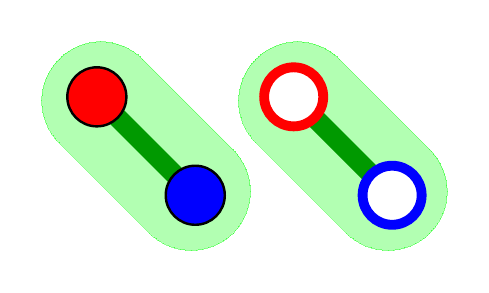}}} \simeq \Sigma_x
= \sum\limits_{k,l} | k,l,R \rangle \langle k, l,B |+ \mathrm{H.c.} \;,
\end{align}
which swaps the red and blue sublattice (see Fig.~\ref{theory:fig:concept}).
In this way, we obtain the Hamiltonian of the pure lattice model specified explicitly in Table~\ref{theory:tab:Ham_RB}.

\begin{table}
\caption{Hamiltonian $H(t)$ of the driving protocol in
``red'' and ``blue'' sublattice representation of the pseudo-spin.}
\vspace*{1ex}
\hspace*{\fill}%
\includegraphics[width=0.95\linewidth]{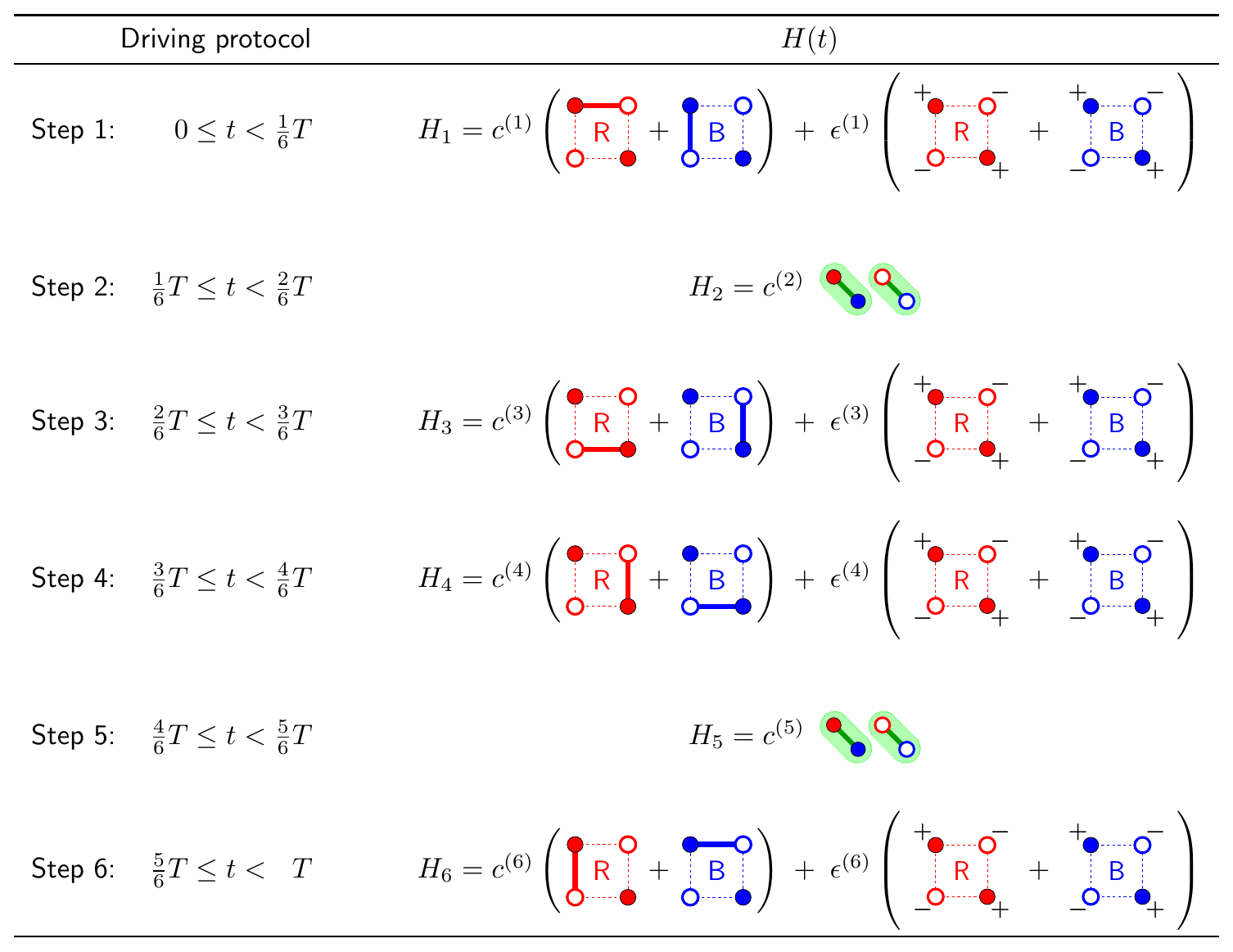}%
\hspace*{\fill}%
\label{theory:tab:Ham_RB}
\end{table}

\subsection{Time-reversal symmetry}

Time-reversal symmetry is defined by the relation
\begin{equation}\label{theory:symm_rel}
 \Theta H(t) \Theta^{-1} = H(T-t)
\end{equation}
(cf. Eq.~(1) in the main text),
where $\Theta$ is an anti-unitary operator with $\Theta^2=1$ for bosonic TRS and $\Theta^2 = -1$ for fermionic TRS.

For fermionic TRS we choose $\Theta = \sigma_y \, \mathcal K$,
with the second Pauli matrix $\sigma_y$ and the operator of complex conjugation $\mathcal K$.
Then, the symmetry relation~\eqref{theory:symm_rel} reads
 \begin{equation}\label{theory:symm_ferm}
\sigma_y  H(t) \sigma_y^{-1} = H(T-t)^*
 \end{equation}
 (and we have $\sigma_y^{-1} = \sigma_y$). Note that the operator $\sigma_y$ only acts on the pseudo-spin degrees of freedom of $H(t)$.

The transformation of terms in the Hamiltonian $H(t)$ is straightforward,
for example, $\sigma_y  \vcenter{\hbox{\includegraphics[width=0.7cm]{model_red1}}}  \sigma_y^{-1} =  \vcenter{\hbox{\includegraphics[width=0.7cm]{model_blue4}}}$, or generally
\begin{align}
\sigma_y |{\uparrow}\rangle\langle{\uparrow}|  \sigma_y^{-1} = |{\downarrow}\rangle\langle{\downarrow}|\;.
\end{align}
On the other hand, we have $\sigma_y  \, \sigma_x \, \sigma_y^{-1} = - \sigma_x$ for
the spin flip $\sigma_x$. Therefore, the driving protocol obeys the relation~\eqref{theory:symm_ferm} if and only if the conditions
 \begin{equation}\label{theory:cond}
 c^{(1)} = c^{(6)}  \;, \quad c^{(3)} = c^{(4)} \;, \quad \epsilon^{(1)} =  \epsilon^{(6)} \;, \quad \epsilon^{(3)} =  \epsilon^{(4)} \;, \quad c^{(2)} = - c^{(5)}
 \end{equation}
are fulfilled.
Then, we have
\begin{equation}\label{theory:trs_hj1}
\Theta H_j \Theta^{-1} = H_{7 - j} \qquad (j = 1, \dots, 6)
\end{equation}
for each of the steps,
or equivalently
\begin{equation}\label{theory:trs_hj2}
\sigma_y H_j \sigma_y^{-1} = H_{7 - j} \qquad (j = 1, \dots, 6)
\end{equation}
since all $H_j$ are real-valued.
 If all parameters are non-zero, the protocol does not possess additional chiral or particle-hole symmetry.

For the present work, we choose the parameters
(spin flip case)
 \begin{equation}\label{theory:params2}
 c^{(1,2,3,4,6)} = 3 \pi/T \;, \quad c^{(5)} = - 3 \pi/T \;, \quad \epsilon^{(1,3,4,6)} = 0 \,,
 \end{equation}
 and (spin rotation case)
 \begin{equation}\label{theory:params1}
 \begin{split}
& c^{(1,3,4,6)} = 5 \pi/(2T) \;, \quad \epsilon^{(1,3,4,6)} = 3/(2T) \;, \\ & \quad c^{(2)} = 2 \pi/T \;, \quad c^{(5)}  = -2 \pi/T \,.
 \end{split}
 \end{equation}

\subsection{Negative coupling}\label{SIsec:negcpl}

The condition~\eqref{theory:cond} implies that either the coupling $c^{(2)}$ in step $2$ or $c^{(5)}$ in step $5$ has to be negative, unless trivially $c^{(2)} = c^{(5)} = 0$.
Negative couplings can indeed be implemented experimentally~\cite{Keil, Kremer}, but we decided to circumvent the additional complexity involved in their implementation and avoid negative couplings. To achieve this, we make the following observation:
In steps 2,5 of the driving protocol, of duration $\delta t$ (here $\delta t = T/6$) and with the spin matrix $c^{(2,5)} \sigma_x$, we have
\begin{align}
 \exp\left( - \mathrm{i} \delta t \, c^{(2,5)} \sigma_x \right) &= \exp \left[ \mathrm{i}n\pi\sigma_x- \mathrm{i} \delta t  \left(\frac{n \pi}{\delta t} + c^{(2,5)} \right) \sigma_x\right]\\
 &= (-1)^n \exp \left[ - \mathrm{i} \delta t  \left(\frac{n \pi}{\delta t} + c^{(2,5)} \right)  \sigma_x\right]
\end{align}
for every $n \in \mathbb Z$.
Therefore, negative couplings $c^{(2,5)} < 0$ in these steps can be replaced by positive couplings $\frac{n \pi}{\delta t} + c^{(2,5)} > 0$ for sufficiently large $n$, without changing the driving protocol implemented in the experiment. For odd $n$, the modified protocol contains an irrelevant global phase.

In the experiment (cf. Methods section), we realize the parameters (spin flip case)
 \begin{equation}\label{theory:params2:pos}
 c^{(1,2,3,4,5,6)} = 3 \pi/T \;,  \quad \epsilon^{(1,3,4,6)} = 0 \,,
 \end{equation}
 and (spin rotation case)
 \begin{equation}\label{theory:params1:pos}
 \begin{split}
& c^{(1,3,4,6)} = 5 \pi/(2T) \;, \quad \epsilon^{(1,3,4,6)} = 3/(2T) \;, \\ & \quad c^{(2)} = 2 \pi/T \;, \quad c^{(5)}  = 4 \pi/T \;,
 \end{split}
 \end{equation}
 having replaced the negative coupling $c^{(5)}$ by the positive value $c^{(5)} + 6 \pi/T$ in step 5 of the driving protocol. Due the global phase introduced by this replacement the Floquet quasi-energies are shifted by $\varepsilon \mapsto \varepsilon + \pi/T$, but the real space propagation remains unchanged.

\subsection{Bulk invariants \& symmetry-protected topological phases}
\label{sec:BulkInvariants}

In order to clearly separate the four topological invariants discussed in the main text, Chern number $\mathcal{C}$, Kane-Mele invariant $\nu_\mathrm{KM}$, Floquet winding number $\mathcal W$ and Floquet TRS invariant $\nu_\mathrm{TR}$, we give an overview of their definition and relevance for (symmetry-protected) topological edge states. For a brief summary, see Tab.~\ref{theory:topoinvariants}.

\begin{table}
\caption{Overview of the discussed topological invariants.}
\vspace*{1ex}
\begin{tabular}{llllll}
\toprule
Invariant & \multicolumn{2}{c}{System type} & Values & Occurence & This work \\\midrule
\rule{0cm}{0.5cm}%
$\mathcal{C}$ & Static & No symmetry & $\mathbb{Z}$ & \cite{Klitzing, Rechtsman} & $\mathcal{C}=0$ \\[3mm]
 $\nu_\mathrm{KM}$ & Static & Fermionic TRS & $\mathbb{Z}_2$ & \cite{Konig2007, Hsieh} & $\nu_{\mathrm{KM}}=0$ \\[3mm]
 $\mathcal W$ & Floquet & No symmetry & $\mathbb{Z}$ & \cite{Maczewsky, Mukherjee} & $\mathcal W=0$\\[3mm]
 $\nu_\mathrm{TR}$ & Floquet & Fermionic TRS & $\mathbb{Z}_2$ & This work & $\nu_{\mathrm{TR}}=1$ \\
 \\[-2.5mm]
\bottomrule
\end{tabular}
\label{theory:topoinvariants}
\end{table}

\subsubsection{Chern number \texorpdfstring{$\mathcal C$}{C}}

The topological classification of time-independent systems without additional symmetries employs the integer-valued Chern number~\cite{TKNN}
\begin{equation}
\mathcal{C}=\frac{1}{2\pi\mathrm{i}}\int_\mathrm{BZ}\mathrm{d}\textbf{k}^2\, \nabla_\textbf{k}\times \langle \psi|\nabla_\textbf{k}|\psi\rangle \; .
\end{equation}
The abbreviation $\mathrm{BZ}$ denotes integration over the entire Brillouin zone. The value of the Chern number corresponds to the net-chirality of edge states.
When evaluated for the individual bands of a Floquet system, the Chern number is calculated from the eigenvectors $\ket{\psi(\mathbf k)}$ of the Floquet-Bloch propagator $U(\mathbf k,T)$.  In Floquet systems, it usually fails to correctly predict the number of edge states~\cite{KitagawaPRB, Lindner} due to the periodicity of the quasi-energy. For the numerical computation of the Chern number, we use the algorithm from Ref.~\cite{Fukui1}.

\subsubsection{Kane-Mele invariant \texorpdfstring{$\nu$}{nu}}

The topological classification of time-independent systems with fermionic TRS employs the $\mathbb Z_2$-valued Kane-Mele invariant~\cite{Fu}
\begin{equation}
\nu_{\mathrm{KM}}=\frac{1}{2\pi\mathrm{i}}\left[\int_\mathrm{BZ_{1/2}}\mathrm{d}\textbf{k}^2\, \nabla_\textbf{k}\times \langle \psi|\nabla_\textbf{k}|\psi\rangle-\int_{\partial \mathrm{BZ}_{1/2}}\mathrm{d}\mathbf k \, \langle \psi|\nabla_\textbf{k}|\psi\rangle\right] \mod 2 \; .
\end{equation}
The abbreviations $\mathrm{BZ}_{1/2}$ or $\partial \mathrm{BZ}_{1/2}$ now denote integration over half of the Brillouin zone or over its boundary, respectively.
A non-zero value of this invariant implies the existence of a pair of symmetry-protected edge states with opposite chirality. Again, when evaluated for the individual bands of a Floquet system, the Kane-Mele invariant is calculated from the eigenvectors of the Floquet-Bloch propagator. Now, symmetry-protected edge states can appear even when the Kane-Mele invariant is zero~\cite{Carpentier, HockendorfPRB, Nathan}, which is indeed the case for our driving protocol.  For the numerical computation of the Kane-Mele invariant, we use the algorithm from Ref.~\cite{Fukui2}.

\subsubsection{Winding Number \texorpdfstring{$\mathcal W$}{W}}
The topological classification of Floquet systems without additional symmetries employs the integer-valued winding number~\cite{Lindner}
\begin{align}
\mathcal W(\varepsilon)=\frac{1}{8\pi^2}\int_0^T \mathrm{d}t \, \int_{\mathrm{BZ}}\mathrm{d}\textbf{k}^2 \, \mathrm{Tr}\left(U_\varepsilon^\dagger \partial_t U_\varepsilon \left[ U_\varepsilon^\dagger \partial_{k_x} U_\varepsilon, \, U_\varepsilon^\dagger \partial_{k_y} U_\varepsilon \right] \right) \; .
\end{align}
This invariant counts the net-chirality of edge states in the band gap at quasi-energy $\varepsilon$. Conceptually, it replaces the Chern number of time-independent systems as the relevant invariant for Floquet systems.

The modified propagator $U_\varepsilon(\mathbf k,t)$ is constructed from the Floquet-Bloch propagator $U(\textbf{k},t)$ as follows:
\begin{align*}
U_\varepsilon(\textbf{k},t)=
\begin{cases}
U(\textbf{k},2t) & \mathrm{if}\quad 0\leq t \leq \frac{T}{2}\\
V_\varepsilon(\textbf{k},2T-2t) & \mathrm{if } \quad \frac{T}{2} < t \leq T\\
\end{cases} \;,
\end{align*}
where $V_\varepsilon(\textbf{k},t)=\exp(t \log_{\varepsilon} U(\mathbf k,T))$. The branch cut of the complex logarithm is chosen along the line from zero to $\mathrm e^{-\mathrm{i} \varepsilon T}$, i.\,e., the eigenvalues of $\log_{\varepsilon} U(\mathbf k,T)$ are elements of the interval 
$\left(T \varepsilon-2\pi,T \varepsilon\right]$.

Alternatively, the winding number $\mathcal W$ may be expressed as the sum
\begin{equation}\label{theory:W_invariant}
\mathcal W(\varepsilon)=\sum_{i=1}^{\mathrm{dp}} N_i(\varepsilon) \hat{\mathcal C}_{i}
\end{equation}
over all degeneracy points $i=1, ..., \mathrm{dp}$ of the Floquet-Bloch propagator $U(\mathbf k,t)$ that occur during time-evolution~\cite{Nathan,HockendorfJPA}. To each degeneracy point, we assign a topological charge $\hat{\mathcal C}_i$, given as a Chern number, and a weight factor $N_i(\varepsilon)$ that ensures that only the degeneracy points in the gap $\varepsilon$ contribute to the sum. Now, the Chern numbers $\hat{\mathcal C}_i$ and weight factors $N_i(\varepsilon)$ are calculated from the eigenvectors and eigenvalues of the Floquet-Bloch propagator $U(\mathbf k,t)$ for all $0 \le t \le T$.
For the numerical evaluation of the $\mathcal W$-invariant, we use the algorithm from Ref.~\cite{HockendorfJPA}.

\subsubsection{TRS invariant \texorpdfstring{$\nu_\mathrm{TR}$}{nuTR}}

In Floquet systems with fermionic TRS, the degeneracy points of the Bloch propagator appear in pairs with opposite topological charge, and cancel each other in the expression for the $\mathcal W$-invariant~\eqref{theory:W_invariant}. The appropriate $\mathbb Z_2$-valued invariant for these systems~\cite{Nathan,HockendorfPRB},
\begin{equation}
\nu_{\mathrm{TR}}(\varepsilon)=\sum_{i=1}^{\mathrm{dp}/2} N_i(\varepsilon) \hat{\mathcal C}_{i} \mod 2 \;,
\end{equation}
counts only one partner of each symmetric pair of degeneracy points, as indicated by the upper summation limit $\mathrm{dp}/2$. A non-zero value of $\nu_{\mathrm{TR}}(\varepsilon)$ implies the existence of symmetry-protected edge states with opposite chirality in the band gap at quasi-energy $\varepsilon$. Conceptually, this invariant serves the same role for Floquet systems as the Kane-Mele invariant for time-independent systems. For the numerical evaluation of the $\nu_{\mathrm{TR}}$-invariant, we use the algorithm from Ref.~\cite{HockendorfJPA}.

\subsection{Topological consideration of a ribbon geometry}

In a finite sample,
symmetry-protected topological phases manifest themselves through chiral edge states.
In our experiment, as well as in the numerical simulations, the edges of the sample run along either $-45^\circ$ (``$x$-axis'') or $+45^\circ$ (``$y$-axis'') on the centered square lattice, as indicated in Fig.~3 and Fig.~\ref{theory:fig:concept}.
Note that the edges have to preserve TRS and thus may not separate lattice sites that are paired in the pseudo-spin representation.

Fig.~\ref{theory:fig:bands} shows the edge states on a semi-infinite ribbon, together with the Floquet bands of the bulk,
using the parameters of our driving protocol in Eq.~\eqref{theory:params2:pos} (spin flip case) or Eq.~\eqref{theory:params1:pos} (spin rotation case).
 In Fig.~\ref{theory:fig:bands}, the ribbon is $15$ unit cells wide, and we only show the edge states on one of the two edges.
Numerically, the edge states and bulk bands are computed from diagonalization of the Floquet propagator 
on the ribbon after one driving period $T$, evaluated as a function of the momentum $k_{x/y}$ parallel to the edges along the $x$-axis or $y$-axis.
Note that we include the shift $\varepsilon \mapsto \varepsilon + \pi/T$ of Floquet quasi-energies that appears through the replacement  $c^{(5)} \mapsto c^{(5)} + 6 \pi/T$ of the negative parameter $c^{(5)}$ by a positive value as we switch from the parameters in Eqs.~\eqref{theory:params1},~\eqref{theory:params2} to the experimental parameters in Eqs.~\eqref{theory:params1:pos},~\eqref{theory:params2:pos} (see Sec.~\ref{SIsec:negcpl}).
Accordingly, the gap appears at quasi-energy $\varepsilon=0$.

Through the bulk-edge correspondence the existence of chiral edge states coincides with a non-zero value of the respective bulk invariants, as collected in Sec.~\ref{sec:BulkInvariants}.
The present situation is characterised by the values listed in Table~\ref{theory:topoinvariants}.
%
Since $\mathcal C=0$ and $\mathcal W=0$ by TRS, edge states have to appear in counter-propagating pairs.
Since $\nu_{\mathrm{KM}} = 0$ but $\nu_\mathrm{TR} \ne 0$ an odd number of counter-propagating pairs of edge states has to be present in the gap between the two Floquet bands.
Note that this combination of invariants corresponds to an anomalous Floquet topological phase~\cite{KitagawaPRB, Lindner}.

Counter-propagating edge states are indeed observed in Fig.~\ref{theory:fig:bands}  (here, a single pair).
In both cases, the edge states exist independently of the direction of the edge, as required for (symmetry-protected) topological states.
In the spin flip case, the Floquet bands are perfectly flat and the dispersion of the edge states is linear.
Changing the parameters of the driving protocol from the spin flip to the spin rotation case, the Floquet bands acquire dispersion but the topological invariants do not change since the gap does not close.
Alternatively, we could note that the number of crossings of the edge state dispersion at the invariant momenta $k_{x,y} = 0, \pi/a$, and hence the number of counter-propagating edge states, is protected by TRS through Kramers degeneracy.
Indeed, these two viewpoints are equivalent due to the bulk-edge correspondence.
The pair of counter-propagating edge states observed here in momentum space gives rise to the propagating modes observed in real space in the experiment (see Figs.~4,~\ref{figsup1},~\ref{figsup2}).

\begin{figure}
\hspace*{\fill}
\includegraphics[scale=0.22]{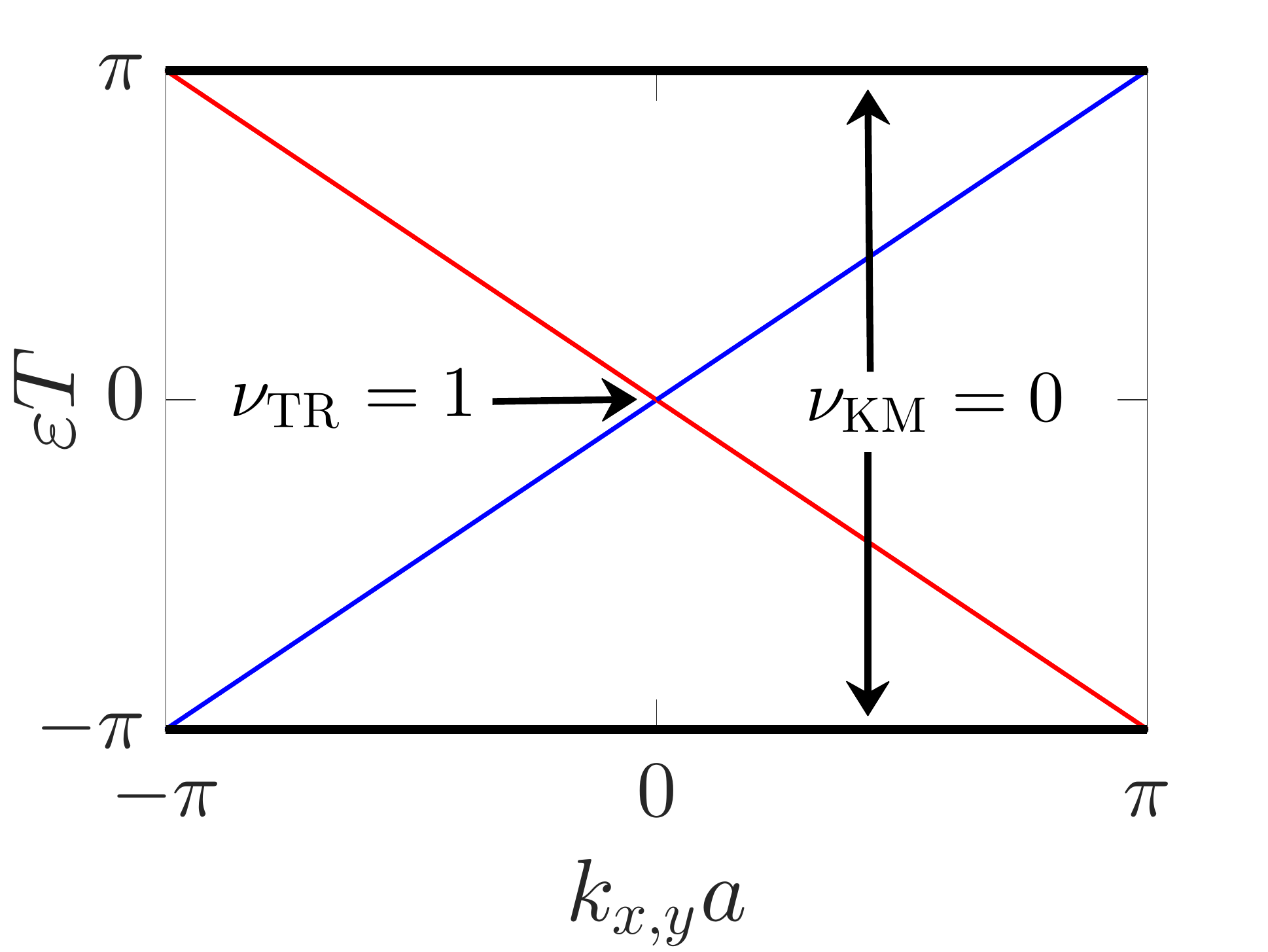}
\hspace*{\fill}
\includegraphics[scale=0.22]{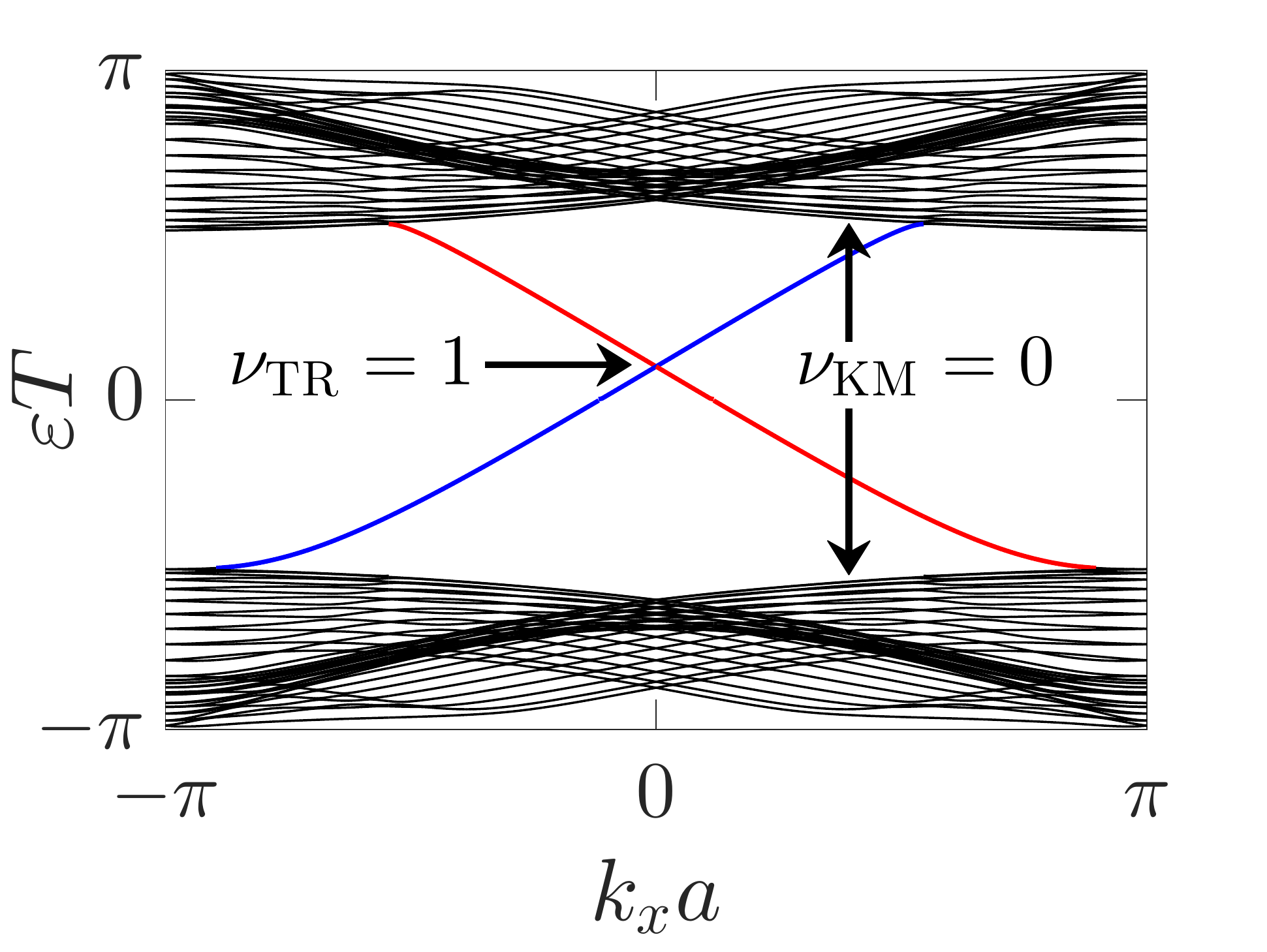}
\hspace*{\fill}
\includegraphics[scale=0.22]{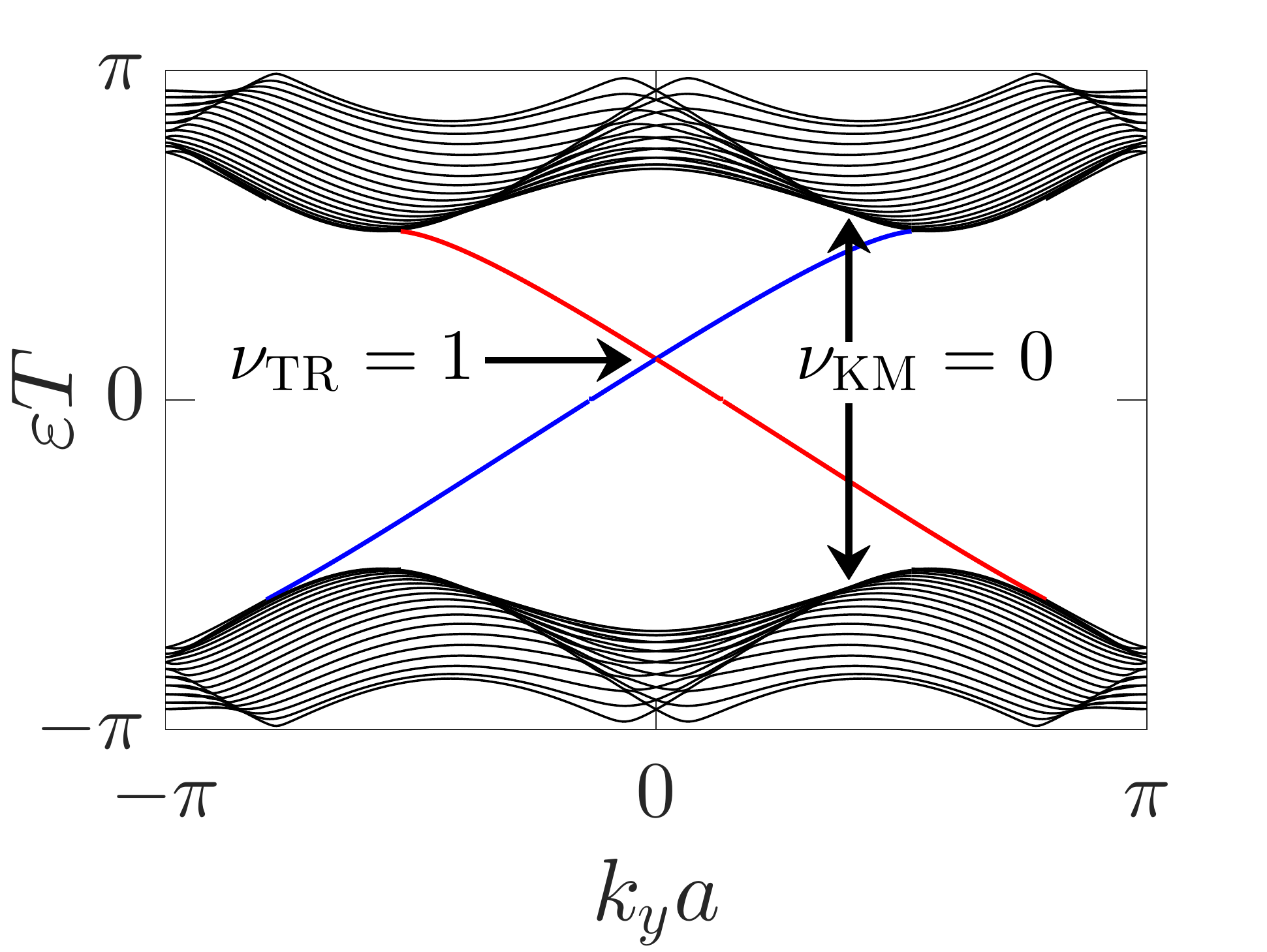}
\hspace*{\fill}
\caption{Floquet bands and symmetry-protected counter-propagating topological edge states for the spin flip (left panel)
and spin rotation (central and right panel) case. Included are the values of the Kane-Mele invariant of the Floquet bands and the $\nu_\mathrm{TR}$-invariant in the central gap.
}
\label{theory:fig:bands}
\end{figure}

\subsection{Probing fermionic time-reversal symmetry}

To check the TRS relation~\eqref{theory:symm_rel} experimentally, we flip the sample as described in the main text.
As we derive now, this allows us to distinguish fermionic from bosonic TRS.

Flipping the sample does not directly correspond to reversing time.
Instead, if the forward propagator is given by Eq.~\eqref{theory:floquet_prop},
the backward propagator of the flipped sample is
\begin{equation}
 \tilde U(T) = U_1 \, U_2 \, U_3 \, U_4 \, U_5 \, U_6 \;,
\end{equation}
as flipping the sample simply reverses the order of steps for the Hermitian Hamiltonians $H_j$.
Here, we consider only one period of the driving protocol. Generalization to several periods is straightforward.

In the present situation, a general TRS operator can be written as $\Theta=\sigma \mathcal{K}$,
with a unitary spin-$\tfrac12$ matrix $\sigma$ such that $\sigma \sigma^* =\pm \mathbbm 1_2$.
For such a general operator, the TRS relation~\eqref{theory:symm_rel} is valid if and only if
\begin{equation}
\sigma H_j \sigma^{-1} = H_{7 - j} \qquad (j = 1, \dots, 6)
\end{equation}
for the Hamiltonians $H_j$ of each step (cf. Eqs.~\eqref{theory:trs_hj1},~\eqref{theory:trs_hj2}). Here, we use that the $H_j$ are real-valued in our driving protocol, which allows us to drop the complex conjugation $\mathcal K$.
Equivalently, we have
\begin{align}
 \sigma U_j \sigma^{-1} &= \exp\big( - \mathrm{i} (T/6) \, \sigma H_j \sigma^{-1} \big) \\
 &=\exp\big( - \mathrm{i} (T/6) \, H_{7-j} \big) = U_{7-j}
\end{align}
for the propagators $U_j = \exp \big(-\mathrm i  H_j T/6 \big)$ of each step.
Therefore, the TRS relation for the backward propagator reads
\begin{align}
\sigma U(T) \sigma^{-1}  = \big(\sigma U_6 \sigma^{-1} \big) \cdots \big(\sigma U_1 \sigma^{-1}  \big) = U_1 \, \cdots \, U_6 = \tilde U(T) \;.
\end{align}
Now suppose we use in the experiment the input state
\begin{equation}
|\psi_\mathrm{in}(\phi)\rangle = |k_0,l_0,\mathrm R\rangle + \mathrm e^{\mathrm i \phi} |k_0,l_0,\mathrm B\rangle \;,
\end{equation}
with finite amplitude on two adjacent red and blue sites and relative phase $\phi$,
which propagates through the unflipped sample, i.e., with forward propagation as in the left panel of Fig.~5(a).
Then, the intensities of the waveguides measured at the output facet are given by the state
\begin{equation}
|\psi_\mathrm{out}(\phi)\rangle= U(T) |\psi_\mathrm{in}(\phi)\rangle  = \sum\limits_{k,l}
\big( \psi_{k,l,\mathrm R}(\phi) |k,l,\mathrm R \rangle +\psi_{k,l,\mathrm B}(\phi) |k,l,\mathrm B \rangle \big)
\;,
\end{equation}
where the amplitudes $\psi_{k,l,\mathrm{R}/\mathrm{B}}(\phi)$ could be computed with the Hamiltonian $H(t)$.
Summing over the red ($\mathrm R$) or blue ($\mathrm B$) sites, respectively, we obtain the output intensities
\begin{equation}
 I^\mathrm{R}(\phi) = \sum\limits_{k,l}  |\psi_{k,l,\mathrm R}(\phi)   |^2 \;,
 \quad
  I^\mathrm{B}(\phi) = \sum\limits_{k,l} |\psi_{k,l,\mathrm B}(\phi)   |^2 \;,
\end{equation}
shown in Fig.~5(a) and Fig.~\ref{fig:OutputIntensities}.

If, alternatively, the input state propagates through the flipped sample, i.e., with backward propagation as in the right panel of Fig.~5(a), the output is given by the state
\begin{equation}\label{theory:psiouttilde}
|\tilde{\psi}_\mathrm{out}(\phi)\rangle= \tilde{U}(T) |\psi_\mathrm{in}(\phi)\rangle  =
\Sigma U(T) \Sigma^{-1} |\psi_\mathrm{in}(\phi)\rangle \;,
\end{equation}
now with different output intensities $\tilde I^\mathrm{R}(\phi)$,
$\tilde I^\mathrm{B}(\phi)$.
The operator $\Sigma$ that appears here is the mapping of the pseudo-spin operator $\sigma$ onto the red and blue sublattice structure of the waveguide implementation (cf. Eq.~\eqref{theory:SigmaX}).
In bra-ket notation, it is
\begin{align}
 \Sigma = \sum_{k,l}   & \big( \sigma_{\uparrow\uparrow}   |k,l,\mathrm R\rangle\langle k,l,\mathrm R| +   \sigma_{\downarrow\uparrow}   |k,l,\mathrm B\rangle\langle k,l,\mathrm R| +  \\ & \sigma_{\uparrow\downarrow}   |k,l,\mathrm R\rangle\langle k,l,\mathrm B| +  \sigma_{\downarrow\downarrow}   |k,l,\mathrm B\rangle\langle k,l,\mathrm B| \big)
\end{align}
for
\begin{equation}
 \sigma = \begin{pmatrix} \sigma_{\uparrow\uparrow}   &  \sigma_{\uparrow\downarrow}  \\
 \sigma_{\downarrow\uparrow}  & \sigma_{\downarrow\downarrow}
 \end{pmatrix} \;.
\end{equation}
From Eq.~\eqref{theory:psiouttilde} we see that the relation between the output intensities $I^\mathrm{R}(\phi)$,
$I^\mathrm{B}(\phi)$ for forward propagation and  $\tilde I^\mathrm{R}(\phi)$,
$\tilde I^\mathrm{B}(\phi)$ for backward propagation depends entirely on the operator $\sigma$ that determines $\Sigma$.
Conversely, if the relation between the output intensities is known from the experiment,
the possible choices of $\sigma$ can be deduced.

\begin{table}
\caption{Relation between output intensities in forward and backward propagation for the four relevant choices of the operator $\sigma$ in the general TRS relation.}
\vspace*{1ex}
\hspace*{\fill}
\includegraphics[scale=1]{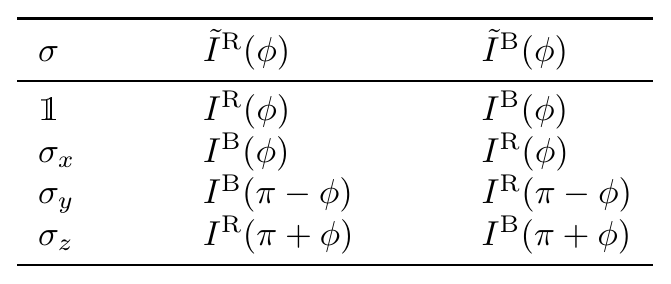}
\hspace*{\fill}
\label{theory:tab:InOut}
\end{table}

The relevant possibilities are listed in Table~\ref{theory:tab:InOut}.
Note that a global phase of the operator $\sigma$ drops out of the TRS relation~\eqref{theory:symm_rel} due to complex conjugation,
and is therefore not included in the table.
For example, with $\sigma \equiv \sigma_y$ we have
\begin{equation}
\Sigma \equiv \Sigma_y = \sum_{k,l}   \big( \mathrm{i}|k,l, \mathrm  B\rangle\langle k,l,\mathrm  R| - \mathrm{i}  |k,l,\mathrm R\rangle\langle k,l,\mathrm B| \big) \;,
\end{equation}
and thus
\begin{equation}
\Sigma_y^{-1} |\psi_\mathrm{in}(\phi)\rangle = -\mathrm{i} \mathrm e^{\mathrm{i} \phi}|\psi_\mathrm{in}(-\phi + \pi)\rangle
\end{equation}
for the input state while, according to Eq.~\eqref{theory:psiouttilde},
\begin{align}
|\tilde{\psi}_\mathrm{out}(\phi)\rangle &=-\mathrm{i} \mathrm e^{\mathrm{i} \phi} \Sigma_y |\psi_\mathrm{out}(-\phi+\pi)\rangle \\
&=
\mathrm e^{\mathrm{i} \phi} \sum\limits_{k,l}
\big( \psi_{k,l,\mathrm R}(-\phi + \pi) |k,l,\mathrm B \rangle - \psi_{k,l,\mathrm B}(- \phi + \pi) |k,l,\mathrm R \rangle \big) \notag
\end{align}
for the output state.
The phases $\pm \mathrm e^{\mathrm{i} \phi}$ drop out, but the output intensities on the red and blue sublattice are swapped by $\Sigma_y$.
Therefore, we get the relations $\tilde I^\mathrm{R}(\phi) = I^\mathrm{B}(-\phi+\pi)$, $\tilde I^\mathrm{B}(\phi) = I^\mathrm{R}(-\phi+\pi)$
given in Table~\ref{theory:tab:InOut}.

Now, the type of TRS realized by the driving protocol
can be determined conclusively from the experimental data in Fig.~5(a) in the main text.
In the experimental data we observe that (i) the output intensities on the red and blue sublattice are swapped
and (ii) a phase shift $\phi \mapsto \pm \phi + \pi$ occurs when flipping the probe.
Observation (i) rules out all possibilities for TRS apart from the choices $\sigma = \sigma_x$ or $\sigma = \sigma_y$,
which are the only operators with purely off-diagonal elements as required for the swapping of intensities.
Observation (ii) rules out all possibilities for TRS apart from the choices $\sigma = \sigma_y$ or $\sigma = \sigma_z$, which are the only operators leading to a phase shift $\phi \mapsto \pm \phi + \pi$.
In combination, we are left with the choice $\sigma = \sigma_y$ of fermionic TRS.

\begin{figure}
	\centering
 	\includegraphics[width=1\textwidth]{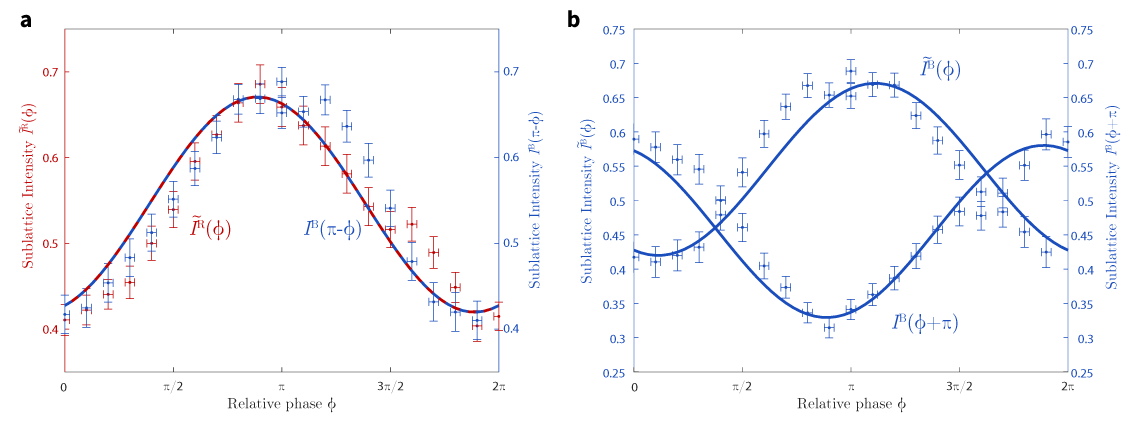}
\caption{
Probing the output intensities from Fig.~5(a) for (a)
fermionic TRS with $\sigma = \sigma_y$ or (b) bosonic TRS with $\sigma = \sigma_z$.
According to Table~\ref{theory:tab:InOut}, it should hold
$\tilde I^\mathrm{R}(\phi) = I^\mathrm{B}(\pi - \phi)$ in case (a) and $\tilde I^\mathrm{B}(\phi) = I^\mathrm{B}(\pi + \phi)$ in case (b) if the respective TRS is realized.
Clearly, the relation for case (a) is satisfied but for case (b) is not.
}
\label{fig:OutputIntensities}
\end{figure}

For a final check of fermionic TRS, the experimental data are reproduced in Fig.~\ref{fig:OutputIntensities} in direct correspondence to the relations from Table~\ref{theory:tab:InOut}.
Note that we have $I^\mathrm{R}(\phi) = 1 - I^\mathrm{B}(\phi)$ and $\tilde I^\mathrm{R}(\phi) = 1 - \tilde I^\mathrm{B}(\phi)$ for the normalized output intensities,
such that the data in Fig.~5(a) fully determine the four functions entering these relations.
Fig.~\ref{fig:OutputIntensities} clearly shows that (only) the choice $\sigma = \sigma_y$ is compatible with the experimental data: Within the limit of experimental uncertainties, we have $\tilde I^\mathrm{R}(\phi) = I^\mathrm{B}(\pi - \phi)$ (hence also $\tilde I^\mathrm{B}(\phi) = I^\mathrm{R}(\pi - \phi)$ for normalized output intensities).
Therefore, probing fermionic TRS results in a positive result:
The experimental data for the output intensities are compatible with --- and only with --- fermionic TRS.


\bibliographystyle{naturemag}

\end{document}